# PhySeqNet: physics-informed fluorescent foci detection for BARseq spatial transcriptomics

Neha Goswami[1], John Hover[1], Renae Galluccio[1], Yichen Wu[1], Huiqing Zhan[1], Xiaoyin Chen[2], Anthony M. Zador[1]*

*[1]Cold Spring Harbor Laboratory, Cold Spring Harbor, NY, USA*

*[2]Allen Institute for Brain Science, Seattle, WA, USA*

*Correspondence: zador@cshl.edu

**Abstract** BARseq enables spatially resolved gene-expression profiling across intact tissue, but its sensitivity depends on accurate detection and decoding of dense fluorescent rolonies. Existing BARseq spot-calling approaches can be conservative in regions with overlapping signal, while supervised deep learning is limited by the lack of reliable manual ground truth in real tissue images. Here we present PhySeqNet, a physics-informed deep learning framework for annotation-free fluorescent foci detection and demonstrate its applicability in spatial transcriptomics (BARseq). PhySeqNet trains U-Net detectors entirely on semi-synthetic images generated from an instrument-matched optical image generation model combined with real sample-free experimental noise, eliminating the need for manual annotation. The detector is integrated into pyBARseq, a modular Python pipeline for BARseq data analysis. On held-out simulated data, PhySeqNet achieved high detection performance across wavelength bands. On real BARseq tissue images, where ground-truth rolony annotations are unavailable, we evaluated detections by forward-model reconstruction of the detected coordinates. PhySeqNet combined with conservative barcode-aware non-maxima suppression (NMSB) achieved higher reconstruction fidelity, measured by normalized cross-correlation (NCC) and multi-scale structural similarity (MS-SSIM), than the single-shot BarDensr density estimator, tuned classical spot detectors, and a pretrained deep-learning spot detector. In the analyzed brain section, PhySeqNet with NMSB yielded a median of 51 counts per cell compared with 47 for BarDensr after quality control, while preserving a similar median number of genes detected per cell. The additional recovered transcripts yielded spatial expression patterns qualitatively consistent with Allen MERFISH reference data and preserved cell-type composition after scANVI label transfer. These results show that physics-informed simulation can train practical, annotation-free detectors for spatial transcriptomics and improve BARseq sensitivity without disrupting downstream biological interpretation.

## Introduction

Spatial transcriptomics is rapidly transforming biomedical research by enabling gene expression to be measured in its native tissue context with approaches such as spatially barcoded transcript-capture methods, multiplexed fluorescence hybridization methods including MERFISH and seqFISH, and image-based in situ sequencing methods like ISS, BARseq, and STARmap [1-12]. Its power comes from its inherently interdisciplinary nature: it combines molecular biology and genomics with optical imaging, image processing, computational modeling, and bioinformatics. As these technologies scale to larger tissue areas, higher spatial resolution, and more complex molecular readouts, the analytical challenge increasingly shifts from data generation to data interpretation [13]. In particular, advances in computer vision and physics-informed computational imaging can play a central role in extracting more accurate molecular information from increasingly complex imaging data [14, 15].

BARseq illustrates this challenge [7, 8]. In BARseq, gene identities are inferred from multi-cycle fluorescence images of rolonies, which can be densely packed, optically blurred, overlapping, and embedded in structured tissue background. The number of decoded transcripts is therefore constrained not only by chemistry and imaging hardware, but also by how effectively the analysis pipeline extracts rolony positions from the measured fluorescence images. Improving rolony detection is thus not simply a matter of reducing image-analysis error; it provides a route to increasing molecular throughput from existing BARseq images by more faithfully decoding the optical image-formation process.

Existing BARseq analysis pipelines rely on BarDensr's single-shot density estimator [16], a fast correlation-based approach that estimates barcode-specific evidence from the multicycle image stack and codebook using a user-defined noise floor. In our data, the single-shot estimator produced comparatively conservative calls in crowded regions, motivating evaluation of a detector trained specifically for source recovery under overlapping fluorescence. Classical spot detectors based on Laplacian-of-Gaussian filtering, Difference-of-Gaussian filtering, or regional/local-maxima detection are fast, interpretable, and widely used for fluorescence spot detection, but their performance depends strongly on scale and intensity-threshold parameters and they are not designed to infer multiple overlapping emitters from a single blended point-spread-function spot [17]. Deep learning offers a natural alternative, but supervised training typically requires manually annotated ground truth [18]. For real BARseq tissue images, such annotations are difficult to obtain at scale and are inherently ambiguous in dense, low-SNR, or overlapping regions with inherent subjectivity arising from inter-operator bias in ground truth annotations [19]. Modern deep-learning spot detectors such as deepBlink [20], Polaris [15], Spotiflow [21], and Piscis [18] have substantially improved fluorescence spot detection across single-molecule FISH, image-based spatial transcriptomics, and general fluorescence microscopy. However, these methods primarily formulate the task as detecting and localizing individual spot-like objects. In contrast, BARseq rolony detection often requires recovering multiple heterogeneous RCA-derived emitters whose fluorescence signals overlap within a shared optical footprint, making the problem closer to source demixing than conventional spot detection.

This motivates an annotation-free training strategy in which known projected emitter locations are generated by a calibrated optical image-formation model rather than manual annotation of experimental tissue images. In fluorescence microscopy, the observed image is determined by the underlying emitter distribution, the optical point-spread function, detector sampling, and the noise statistics of the instrument and tissue. If these components can be modeled realistically, large amounts of training data with exact ground-truth coordinates can be generated synthetically. Simulated training and benchmark data have previously been used in deep-learning approaches for single-molecule localization microscopy and fluorescence spot detection, including DECODE, deepBlink, Spotiflow, Deep-STORM and Piscis [18, 20-23]. This work extends this strategy to spatial transcriptomics (BARSeq) avoiding manual annotations completely.

Here we introduce PhySeqNet, a physics-informed deep-learning detector for fluorescent emitter detection that requires no manual annotation. Here, "physics-informed" refers to the incorporation of physical knowledge through the training data, corresponding to an observational bias in physics-informed machine-learning taxonomies [24, 25]. The network architecture and loss are standard, while the training images are generated using a calibrated optical forward model with known emitter labels. PhySeqNet is trained entirely on semi-synthetic images generated by a calibrated Born and Wolf-model PSF [26] (as applied to a confocal microscope) convolved with realistic emitter volumes and composited with real experimental noise frames, the latter serving as a domain-grounding strategy that captures a realistic distribution of detector noise and optical artifacts without parametric approximation. We couple PhySeqNet with argmax basecalling into an end-to-end detection-and-decoding module.

We also present pyBARseq, a modular Python implementation of the BARseq analysis workflow. pyBARseq supports preprocessing, denoising, background subtraction, channel alignment, bleedthrough correction, cycle registration, stitching, segmentation, spot detection, basecalling, quality control, and generation of AnnData-compatible outputs. An accompanying napari viewer enables interactive inspection of raw images, detection overlays, segmentation masks, basecalls, and spatial gene-expression maps.

We validate PhySeqNet using a forward-model image reconstruction benchmark adapted from the SQUIRREL framework [27], demonstrating higher detection fidelity than single shot density estimator from BarDensr [16], classical spot detectors, and the pretrained deep-learning detector Piscis [18]. We show that PhySeqNet's highly sensitive transcript recovery preserves biological structure, assessed by qualitative comparison of spatial expression patterns with an independent MERFISH reference and cell-type classification via scANVI label transfer [28, 29] from the Allen Brain Cell Atlas which integrates whole-brain single-cell RNA-seq and MERFISH atlases of the mouse brain [30, 31].

## Results

### pyBARseq: an open-source Python pipeline for BARseq spatial transcriptomics

The existing BARseq computational pipeline relies on MATLAB which makes it difficult to integrate with modern Python-based image processing libraries and packages [8]. To address these limitations, we developed pyBARseq-(an adaptation of the original MATLAB pipeline [8]), a modular, end-to-end Python pipeline that integrates all stages of BARseq data processing into a single configurable framework (fig. 1).

pyBARseq accepts raw multi-cycle, multi-channel BARseq imaging data and produces analysis-ready AnnData objects containing per-cell expression matrices, spatial coordinates, barcode identities, and slice assignments enabling direct integration with the SCANPY ecosystem [32] for downstream clustering, label transfer, and visualization.

To facilitate interactive quality control and data exploration, we developed a companion napari plugin (pybarseq_viewer, Supplementary Fig. S13) that provides a six-panel tabbed dock widget for simultaneous visualization of raw fluorescence images, detection overlays, cell segmentation masks, basecalling results, and spatial gene expression maps along with some quantitative metric inspection to determine tile usability. This viewer enables rapid, tile-by-tile inspection of each processing stage, supporting both method development and routine quality assessment.

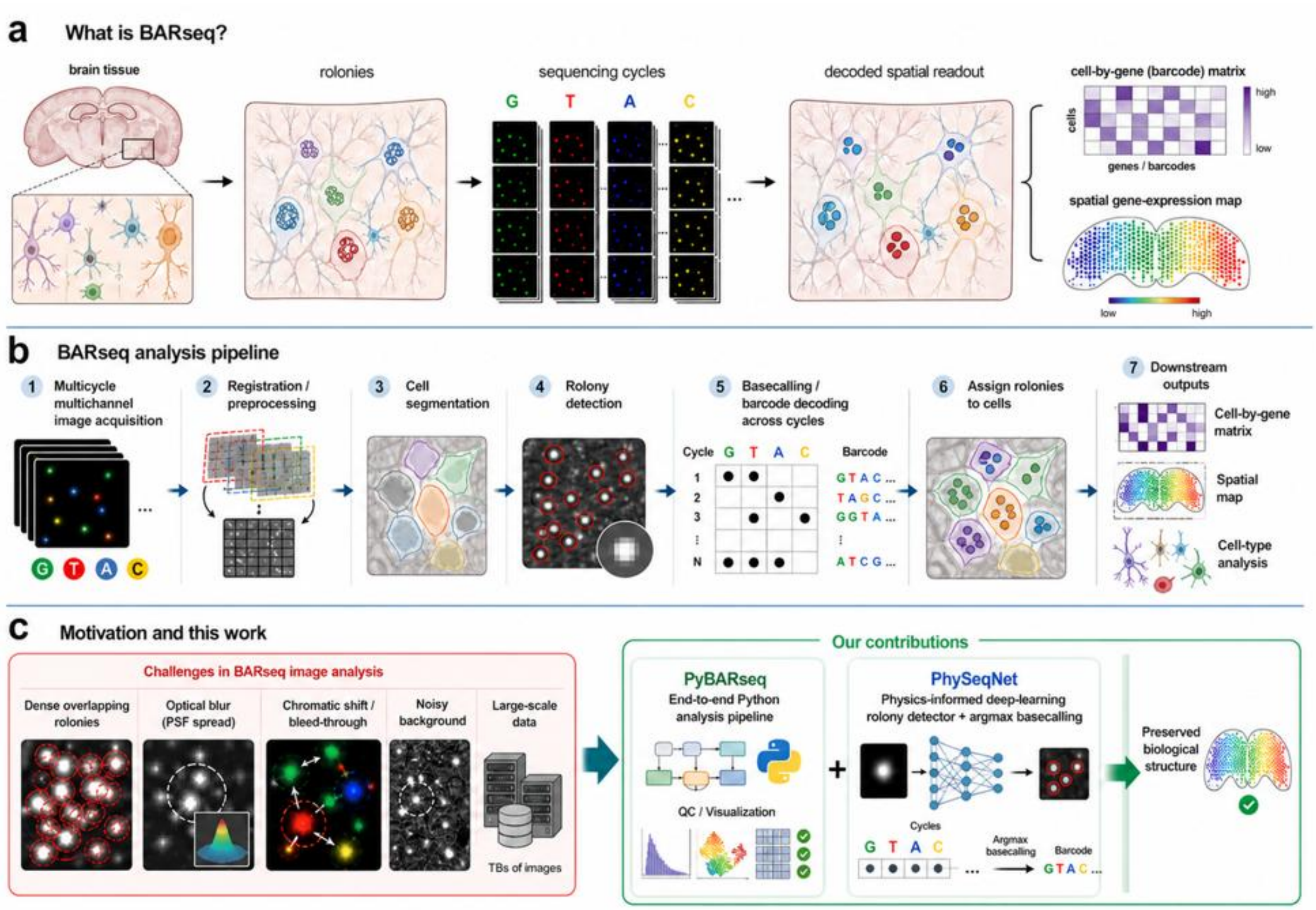

*Figure 1. pyBARseq: a modular Python pipeline for end-to-end BARseq spatial transcriptomics analysis. a. Schematic overview of BARseq, in which rolonies are read out across multiple sequencing cycles and decoded into spatially resolved gene-expression measurements. b. Overview of the pyBARseq analysis workflow, from multicycle image acquisition and preprocessing through cell segmentation, rolony detection, basecalling, rolony-to-cell assignment, and downstream cell-by-gene and spatial analyses. c. Summary of the main analytical challenges in BARseq image analysis and the contributions of this work: pyBARseq provides an end-to-end Python analysis and QC framework, while PhySeqNet combines physics-informed rolony detection with argmax basecalling while preserving biological structure.*

## Physics-informed simulation enables annotation-free training for rolony detection

A major bottleneck in applying deep learning to rolony detection is the need for manually annotated training data that can introduce operator to operator bias in the ground truth. Because ground-truth rolony positions are not available in real BARseq images, we developed a physics-informed simulation pipeline that generates realistic semi-synthetic training data directly from the instrument's optical model and measured noise, eliminating the need for any manual annotation (fig. 2). The Born and Wolf model for the optical PSF was used [26]. For wavelength $\lambda$, objective numerical aperture NA, immersion refractive index $n_i = 1$, and wavenumber $k = 2\pi n_i/\lambda$ in the immersion medium, the PSF intensity at lateral coordinate $\mathbf{r}_\perp$ and axial coordinate $z$ was computed, up to a constant amplitude factor, using the paraxial approximation following Ref. [26] as

$$h(\mathbf{r}_\perp, z) = \left| \int_0^1 J_0 \left( k \frac{\mathrm{NA}}{n_i} |\mathbf{r}_\perp| \rho \right) \exp\left( -i \frac{k}{2} \rho^2 z \frac{\mathrm{NA}^2}{n_i^2} \right) \rho \, d\rho \right|^2,$$

where $J_0$ is the zeroth-order Bessel function of the first kind; the integral over $\rho$ was evaluated numerically via the trapezoidal rule on a uniform $\rho$ grid of 500 points. Excitation and emission PSFs were computed separately at each channel's $\lambda_{\mathrm{exc}}$ and $\lambda_{\mathrm{em}}$ (520/550, 540/590, 640/690, 640/730 nm for G, T, A, C detection channels) and the resulting per channel confocal PSF can be expressed as [33]

$$h_{\mathrm{tot}}(\mathbf{r}) = h_{\mathrm{exc}}(\mathbf{r}) \cdot \left( h_{\mathrm{em}}(\mathbf{r}) *_{xy} A(\mathbf{r}_\perp) \right),$$

where $A$ is the pinhole aperture function. To incorporate the anisotropy introduced in the PSF shape due to aberrations, we calibrate the spatial anisotropy using real fluorescent beads images and incorporate the anisotropy in our simulated PSF (Supplementary fig. S1).

A 3D volume $V_{\mathrm{em}}(\mathbf{r})$ of emitters signifying both inter and intra-cytoplasmic emitters is modelled based on the measured size distribution of real rolonies (Methods Section). The measured (noiseless) image volume can then be expressed as a convolution of the true emitter volume and the confocal PSF.

$$V_{\mathrm{meas}}(\mathbf{r}) = V_{\mathrm{em}}(\mathbf{r}) \circledast h_{\mathrm{tot}}(\mathbf{r})$$

where $\circledast$ denotes 3D convolution.

For BARseq, we operate on maximum intensity projected images per volume, so this volume is also transformed into a maximum intensity projected image denoted by $I_{sim}$. To bridge the simulation-to-real gap, sample-free experimental images $I_{empty}$ were used directly as the noise source rather than synthesizing noise from a parametric model. These frames capture instrument- and acquisition-specific background components, including camera read noise, dark current, fixed-pattern noise, and optical background, without requiring an explicit parametric noise approximation. The resulting semi-synthetic image is then

$$I_{\mathrm{train}} = \mathrm{cast}_{\mathrm{uint16}}\left( I_{\mathrm{sim}} + I_{\mathrm{empty}} \right).$$

Known centroids of the emitters are then taken as the ground truth map (fig. 2). Implementation details of the simulation are mentioned in the Methods Section.

**PhySeqNet detection generalizes across fluorescence channels with high precision and recall**.

We trained two wavelength-band specific U-Net models (Supplementary fig. S2), Model G for the green-orange channels (G and T) and Model C for the red to deep red channels (A and C) and evaluated their detection performance on held-out simulated test data (fig. 3).

On held-out simulated data (nonblank images), Model G achieved mean precision = 0.921, recall = 0.879, FDR = 0.079, and miss rate = 0.121 (fig. 3a). Model C, performed comparably with mean precision = 0.934, recall = 0.882, FDR = 0.066, and miss rate = 0.118 (fig. 3b).

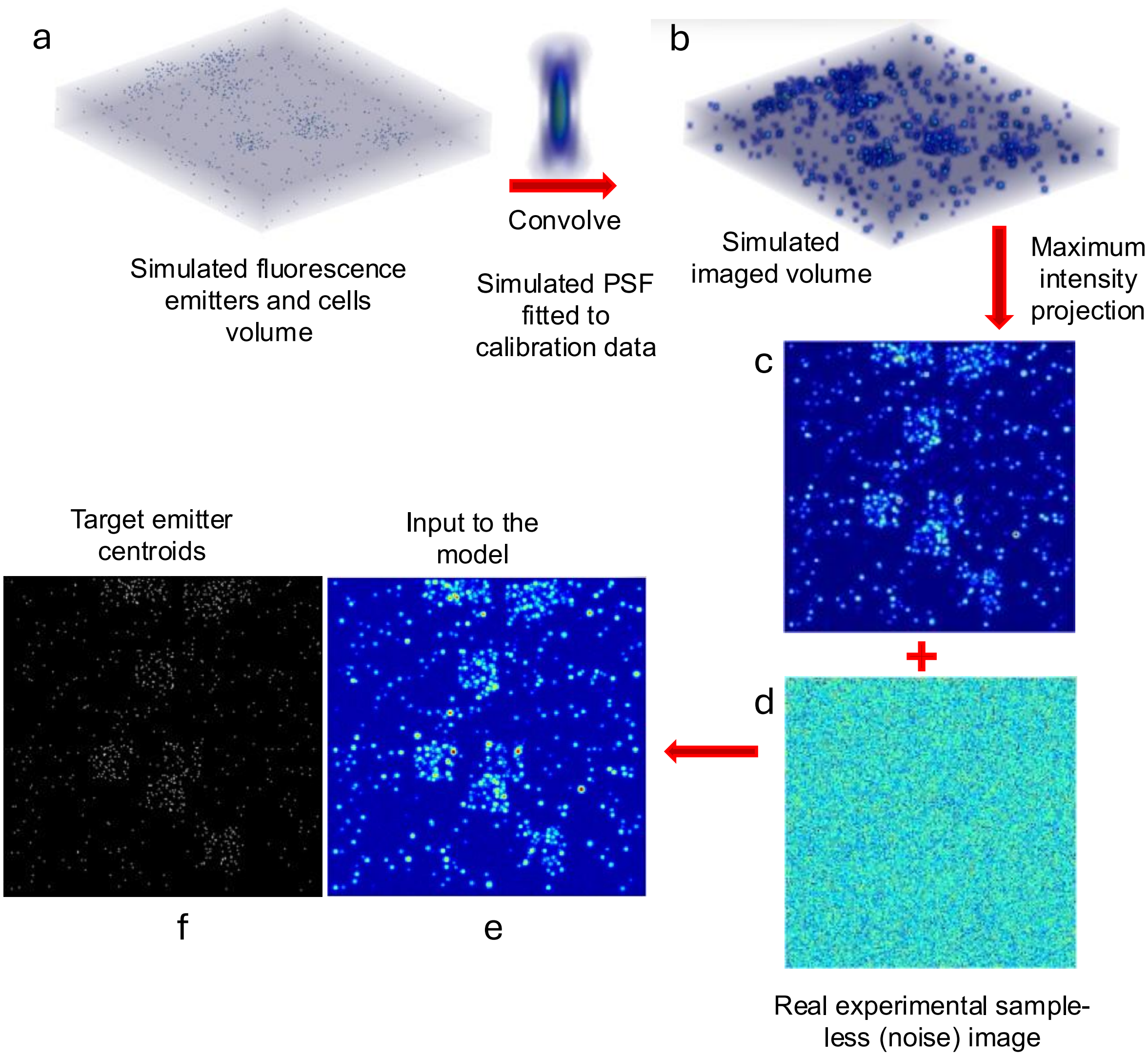

*Figure 2. Physics-informed simulation generates annotation-free training data with exact emitter coordinates and realistic experimental noise. a. Simulated 3D fluorescent emitter volumes were generated to mimic isolated and clustered BARseq rolonies, then convolved with a calibrated confocal PSF. b. The resulting simulated imaged volume was maximum-intensity projected to match the BARseq image representation. c. The projected synthetic signal was combined with d. real sample-free experimental noise images to produce e. semi-synthetic training inputs. f. The known simulated emitter centroids provided binary detection targets, enabling U-Net training without manual annotation of real tissue images.*

The models were then tested on real experimental data. Representative detection overlays across all four channels (figs. 3c-d) show localization of discrete emitter candidates even in regions of high spatial density where classical methods struggle to resolve closely spaced spots.

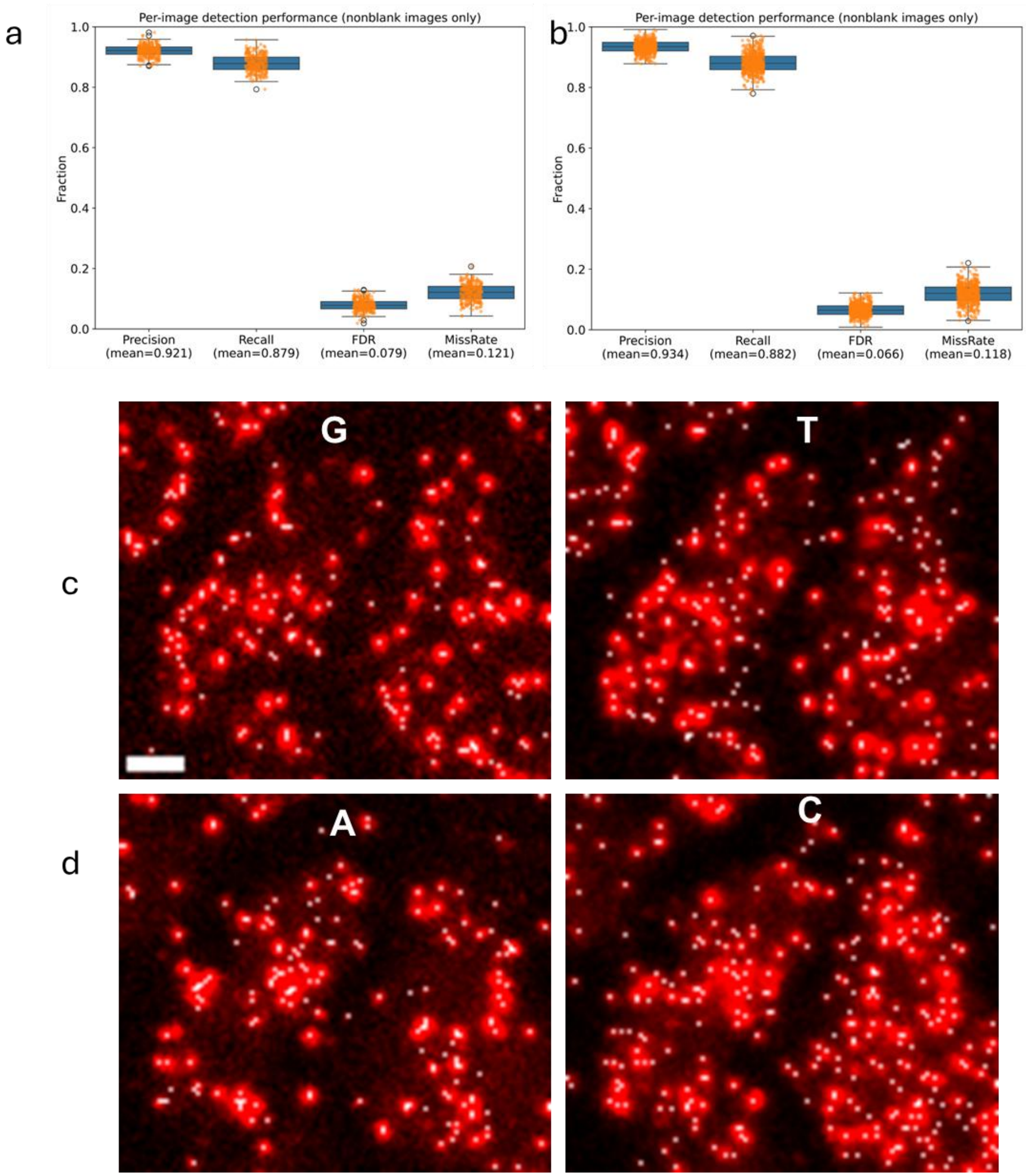

*Figure 3. PhySeqNet achieves high simulated detection performance and generalizes to real BARseq fluorescence channels. a,b. Detection performance on held-out nonblank simulated test images for Model G and Model C, respectively, showing high precision and recall with low FDR and miss rate. c,d. Representative real BARseq image overlays across the G, T, A, and C sequencing channels show that PhySeqNet localizes discrete emitter candidates in dense fluorescence regions. Scalebar 5 μm.*

### Conservative barcode aware non-maxima suppression

PhySeqNet's higher sensitivity can produce islands of emitter locations in crowded regions in the image. This is a property by design where we trained the model to favor dense predictions (more in Methods Section). Consequently, raw PhySeqNet detections can include multiple detections corresponding to the same underlying rolony. To collapse these likely duplicates, we applied barcode-aware non-maximum suppression (NMSB): detections separated by <2 pixels (<0.66 μm) were collapsed to the detection with the higher network score, but only when both detections were assigned the same barcode (see basecalling details in the Methods Section). Detections with different barcodes were retained. The effect of this operation is illustrated in Supplementary Fig. S6, where raw PhySeqNet detections are reduced to a more conservative set of candidate rolonies after NMSB.

### Forward-model reconstruction shows higher detection fidelity on real tissue data

To evaluate detection quality on real BARseq tissue images where no ground-truth annotation exists, we employed a forward-model reconstruction benchmark motivated by SQUIRREL framework [27]: for each method, detected spot coordinates weighed by the original intensity value were convolved with a simplified PSF to reconstruct a synthetic image.

We benchmarked our method raw PhySeqNet and PhySeqNet with NMSB against three classes: conventional spot detectors (Laplacian of Gaussian (LoG), Difference of Gaussian (DoG), h-maxima regional maxima detection), the currently used single-shot density estimator (BarDensr) [16], and deep-learning based spot detector (Piscis) [18] (fig. 4a-g). An important distinction for BarDensr and PhySeqNet is that these are full density estimation/detection respectively plus basecalling methods-meaning their detections are filtered upon basecalling removing spurious uncalled spots.

Across 10 representative fields of view, PhySeqNet and PhySeqNet with NMSB achieved the highest reconstruction fidelity by both NCC and MS-SSIM [34] (Fig. 4). Mean NCC was 0.862± 0.024, 0.848±0.018. MS-SSIM mean values are 0.702 ± 0.026 and 0.706 ± 0.025 respectively. Pairwise comparisons across the 10 fields of view were performed using one-sided paired Wilcoxon signed-rank tests, with Benjamini-Hochberg correction for six comparisons within each metric. For NCC, PhySeqNet with NMSB significantly outperformed all external comparison methods (adjusted $p < 0.05$) but not raw PhySeqNet, which achieved a slightly higher mean NCC. For MS-SSIM, PhySeqNet with NMSB significantly outperformed all comparison methods, including raw PhySeqNet (adjusted $p < 0.05$).

Pairwise directional colocalization analysis, defined as the fraction of detections from method A lying within 2 pixels of a detection from method B, revealed a marked asymmetry between PhySeqNet and conventional spot detectors (Fig. 4g, shown for one field of view). Most LoG, DoG, h-maxima, and Piscis detections colocalized with PhySeqNet (0.93–0.97), whereas only 0.51–0.59 of raw PhySeqNet detections colocalized with these methods, indicating that the conventional detectors recover a largely shared subset of the broader PhySeqNet detection set. In contrast, PhySeqNet and BarDensr showed substantially more reciprocal overlap (0.88 of PhySeqNet detections colocalized with BarDensr and 0.83 of BarDensr detections colocalized with PhySeqNet). Fig. 4e-g also shows an interesting agreement between the classes of methods: spot detectors (LoG, DoG, h-maxima) both conventional and deep-learning based (Piscis) agree well with each other, while density estimator (single shot BarDensr) and PhySeqNet agree more among themselves demonstrating a fundamental difference in operating regime.

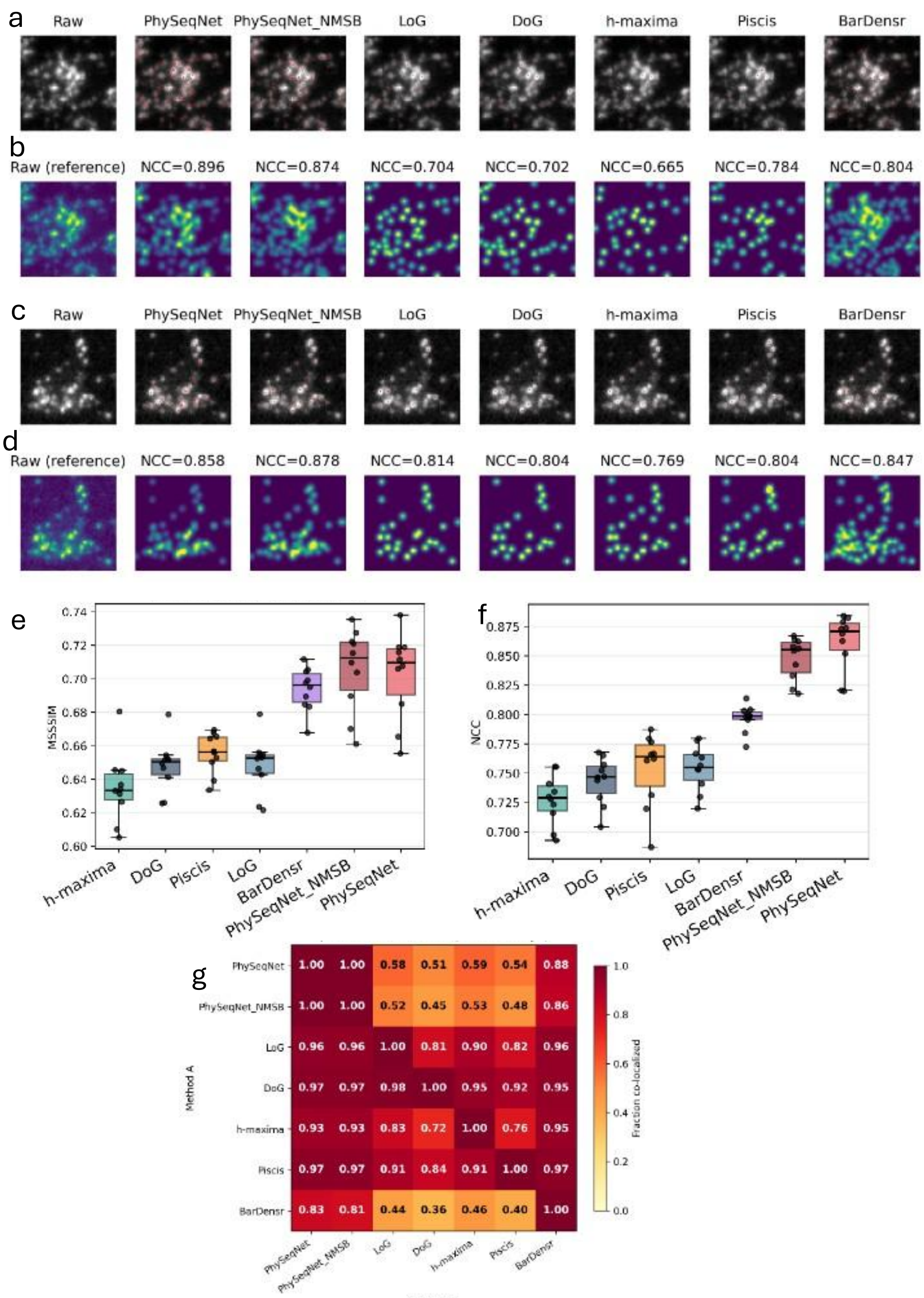

*Figure 4. Forward-model reconstruction demonstrates high-fidelity source recovery by PhySeqNet on real BARseq images. a,c, Representative dense and relatively sparse BARseq image regions, respectively, showing detections from raw PhySeqNet, PhySeqNet with NMSB, classical spot detectors (LoG, DoG, and h-maxima), Piscis, and the single-shot BarDensr estimator overlaid on the measured fluorescence image. b,d, Corresponding forward-model reconstructions. NCC values quantify agreement between each reconstruction and the measured image. e,f, Aggregate distributions of MS-SSIM and NCC across 10 fields of view. Points indicate individual fields of view and boxplots summarize the corresponding distributions. g, Pairwise directional colocalization of detections within a 2-pixel radius for a representative field of view.*

The comparison to Piscis is particularly informative: Piscis, a state-of-the-art deep-learning spot detector [18] trained on a diverse corpus of manually annotated RNA FISH images supplemented with synthetic data, achieved NCC comparable to the tuned classical baselines (0.754) but substantially below PhySeqNet (0.862) when applied zero-shot to BARseq data. This performance gap likely reflects the broader challenge of domain transfer across imaging modalities. In contrast, PhySeqNet's simulation-based training paradigm generates training data using an instrument-matched PSF model, real sample-free experimental noise, and BARseq-like spot densities, allowing the detector to learn features tailored to the target modality without requiring manual annotation.

PhySeqNet's output can also be understood as a sparse representation of the underlying object geometry that leads to faithful forward-model image reconstruction. Importantly, PhySeqNet performs detection, identifying candidate individual emitter positions, rather than foreground segmentation. As a control for the reconstruction metric, dense Otsu foreground segmentation achieved slightly higher NCC despite failing to represent discrete emitter positions, whereas PhySeqNet showed higher MS-SSIM across evaluated fields (Supplementary Fig. S3), illustrating the importance of assessing structural similarity in addition to pixel-wise correlation.

**Out-of-distribution Piscis/smFISH benchmarks define the scope of the trained detector**

We evaluated PhySeqNet's Model G (trained on symmetric PSF simulations, offering the broadest generalization potential) on selected Piscis benchmark smRNA FISH images as an out-of-distribution stress test (Supplementary figs. S7-S9) [18]. These datasets differ from BARseq in imaging modality (widefield vs. confocal), objective magnification, numerical aperture, noise characteristics, background structure, spot morphology, and annotation convention. These images are drawn from the Piscis benchmark test datasets, providing a comparison between Piscis evaluated within its native domain and PhySeqNet applied far outside its training distribution. On favorable images (Supplementary fig. S7) (those acquired at comparable wavelengths) with moderate spot density and adequate signal-to-noise ratio, PhySeqNet recovered the majority of annotated spots (F1 = 0.68-0.91), suggesting that the learned source-localization behavior transfers partially to diffraction-limited spot detection in widefield microscopy. The F1 scores are driven down primarily by false positives rather than missed detections, indicating that the core source-localization capability transfers well across modalities even when the background rejection does not. The Piscis annotations deliberately excluded ambiguous low-SNR regions where annotators could not reach a definitive judgment, meaning that a fraction of the apparent false positives may correspond to real spots below the annotation confidence threshold. Visual inspection of the false positive detections (Supplementary fig. S9) reveals that many correspond to faint fluorescent puncta that are plausibly real spots excluded from the ground truth under the Piscis annotation convention of omitting ambiguous low-SNR regions. PhySeqNet achieved recall of 99.5% (426/428) and 95.6% (303/317) on the representative images from Supplementary fig. S7, with only 2 and 14 false negatives respectively, indicating that the simulation-trained detector can generalize as a source localizer even across imaging modalities. The 86 and 269 false positives respectively (which lower the F1) may therefore overestimate the true false positive rate. Performance, however, degraded on images with high diffuse background fluorescence and very sparse spots (F1 = 0.76, but visible false positives in larger spots) and failed on near-empty fields dominated by background structure (F1 = 0.00), where the model's prior becomes inappropriate under widefield background statistics (Supplementary fig. S8). By comparison, Piscis achieved F1 scores of 0.8–0.97 on these same images within its native test distribution. This asymmetry is expected and informative: PhySeqNet was designed for BARseq-like confocal imaging, and while the simulation-trained features transfer promisingly to optically similar conditions, robust deployment on other smFISH modalities would require retraining the simulation pipeline with matched optical parameters and background/noise conditions which is a straightforward extension of the physics-informed training framework that requires no manual annotation.

**High-resolution imaging provides an orthogonal validation of PhySeqNet detections**

To provide an orthogonal validation independent of the forward-model reconstruction benchmark, the same tissue region was re-imaged on a Zeiss Airyscan system using a 63x/1.4 NA objective. After registration and resampling to a common coordinate system, spots in the high-resolution image were detected using a Laplacian-of-Gaussian detector with its threshold selected from the high-resolution image alone. PhySeqNet detections obtained from the original lower-resolution BARseq image showed strong spatial concordance with these independent high-resolution reference detections (Supplementary Fig. S5). PhySeqNet also recovered the dominant punctate structures when applied to the resampled Airyscan image, despite the change in microscope, and image statistics, providing an additional test of source-localization robustness outside the acquisition conditions used for training.

**Transcript recovery preserves gene-level accuracy and spatial expression patterns**

For the rest of the analysis, we compare PhySeqNet with NMSB to its predecessor single-shot density estimator from BarDensr which is the currently used method for BARseq analysis [16]. To assess whether the PhySeqNet's detections translate to biologically valid transcript recovery, we compared per-cell expression statistics and spatial gene expression patterns between the PhySeqNet and BarDensr pipelines (fig. 5).

As shown in fig. 5a, across 60 field of views from the analyzed brain section, per-gene mean expression (across non-zero cells) was systematically higher for PhySeqNet (with and without NMSB), with majority of points lying consistently above the diagonal in a BarDensr-versus-PhySeqNet scatter plot with NMSB bringing the points towards the diagonal. After quality control (cells with ≥5 genes and ≥20 total counts retained), PhySeqNet with NMSB recovered a median of 51.0 counts per cell compared to 47.0 for BarDensr (fig. 5b), a ~8.5% increase. The estimated FDRs (Supplementary Note 8) were 4.8% for raw PhySeqNet, 5.7% after NMSB, and 4.8% for BarDensr. The slight increase in unused-barcode FDR after NMSB likely reflects preferential collapsing of spatially clustered detections from genuinely expressed barcodes, which are far more abundant than unused-barcode calls. Importantly, the number of genes detected per cell was similar between the methods (median = 26, 25) (fig. 5c). An important thing to note is that the raw PhySeqNet counts were much higher (68.0) and were conservatively collapsed by NMSB. NMSB is crucial for reducing false positive rolony detections but on the simulated test dataset it was shown to remove a significant amount of true positive detections also (Supplementary Fig. S4, true positives removed ~40% while removing ~66% false positives, where all detections were assumed to be same barcodes in multipixel components). The post-NMSB transcript count therefore represents an intentionally conservative estimate rather than the maximal sensitivity of the underlying detector as can also be seen in Supplementary Fig S6.

Spatial expression maps for two representative genes, C1ql3 and Lamp5, are shown in Fig. 5d together with Allen MERFISH data from a comparable coronal section (imputed expression) [30]. PhySeqNet with NMSB preserved the broad anatomical expression patterns observed in the independent MERFISH reference for both genes.

### Cell-type composition is concordant between pipelines and consistent with the Allen Brain Cell Atlas

To assess whether the transcript recovery from PhySeqNet affects cell-type identification, we performed label transfer from the Allen Brain Cell Atlas single-cell RNA-seq reference [30] using scVI/scANVI [29] (Methods Section). PhySeqNet with NMSB recovered class- and subclass-level spatial organization broadly consistent with an independent reference Allen MERFISH coronal section (Fig. 6a–d) [30]. In complementary comparisons with BarDensr, class-level cell-type composition remained highly concordant between pipelines (Pearson r=0.993, Spearman r=0.976; Supplementary Fig. S11). Subclass-level composition was similarly preserved (Pearson r=0.990, Spearman r=0.964; Supplementary Fig. S12), although prediction confidence was lower at the subclass level. The distribution of scANVI prediction confidence was similar between processing pipelines (Supplementary Figs. S10–S12).

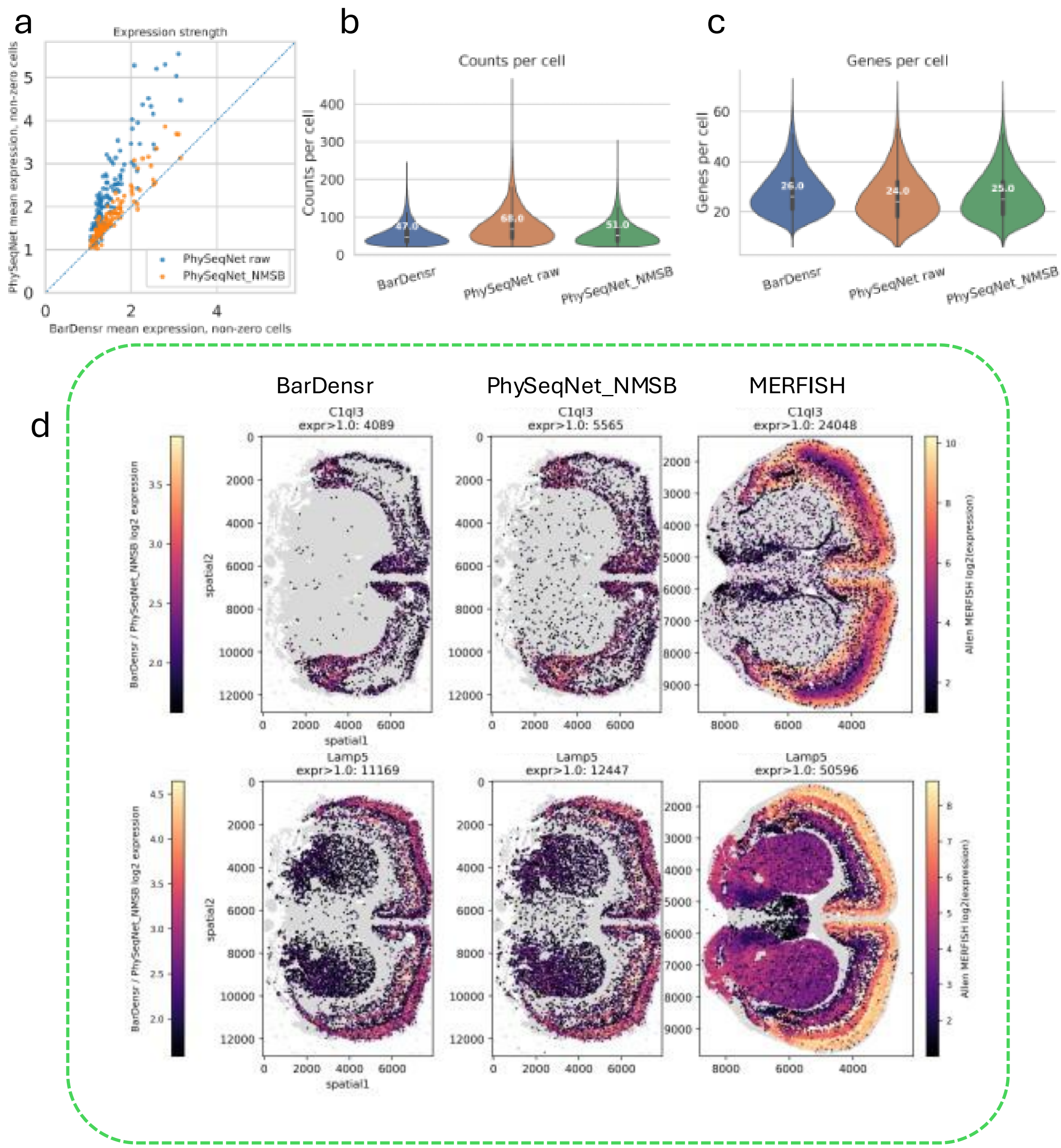

*Figure 5. PhySeqNet increases transcript recovery while preserving gene-level and spatial expression patterns. a, Per-gene mean expression across nonzero cells for the 106-gene sequencing panel, comparing BarDensr with raw PhySeqNet and PhySeqNet with NMSB across 60 fields of view from the analyzed brain section. b, Distribution of counts per cell after cell-level quality control for BarDensr, raw PhySeqNet, and PhySeqNet with NMSB. Median counts per cell were 47, 68, and 51, respectively. c, Distribution of genes detected per cell for the three processing methods. d, Representative spatial expression maps for C1ql3 and Lamp5 obtained with BarDensr and PhySeqNet with NMSB, shown alongside independently acquired Allen Brain Cell Atlas MERFISH expression patterns from a comparable coronal section [30].*

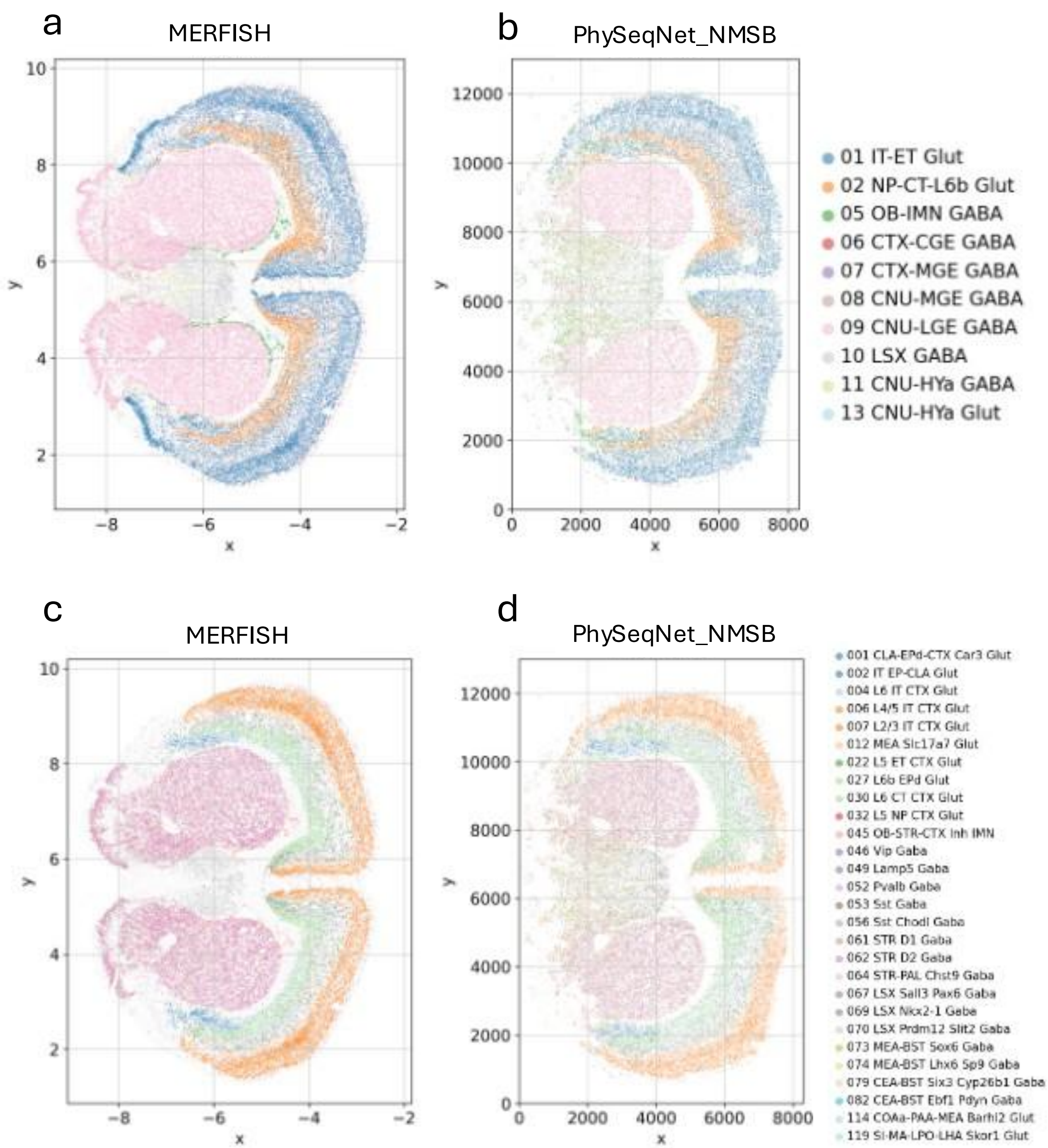

*Figure 6. PhySeqNet with NMSB recovers anatomically coherent cell-type organization consistent with the Allen Brain Cell Atlas [30]. Cell identities in BARseq data were assigned by scVI/scANVI label transfer from the Allen Brain Cell Atlas single-cell RNA-seq reference and compared with an Allen MERFISH coronal section. a,b, Spatial distributions of major cell classes in the Allen MERFISH reference and PhySeqNet-processed BARseq section, respectively. c,d, Corresponding spatial distributions at the subclass level.*

## Discussion

We have presented PhySeqNet, a physics-informed deep learning approach to fluorescent foci detection in BARseq spatial transcriptomics. By training a U-Net entirely on synthetic data generated through Born and Wolf model PSF simulation with real experimental noise, PhySeqNet eliminates the need for manual annotation while achieving higher detection performance compared to commonly used existing methods. We used sample-free experimental frames as the noise source, rather than a parametric noise model, because fixed-pattern noise and instrument-specific background are difficult to capture analytically. This choice, together with the calibrated PSF, was sufficient for simulation-trained detectors to transfer to real tissue images; a direct comparison with parametric noise models was beyond the scope of this work.

The use of NCC and MSSSIM on image reconstruction as an evaluation metric addresses a fundamental challenge in spatial transcriptomics: the absence of ground truth for foci detection in real tissue. While simulation benchmarks provide precision and recall estimates, these cannot fully capture the complexity of real biological samples. The reconstruction quality offers a complementary, ground-truth-free assessment that can be computed on any real image without reference annotations.

The concordance of PhySeqNet derived spatial gene-expression patterns with MERFISH [5, 30, 31], an independent technology based on entirely different biochemistry (sequential hybridization rather than in situ sequencing), provides complementary evidence that the recovered transcripts preserve biological spatial organization.

Two limitations of this work should be noted. First, our validation is performed on a single coronal brain section from one animal. While the simulation-trained detector is in principle generalizable, the full pipeline including cell typing and MERFISH validation has been demonstrated only on this dataset. Second, the out-of-distribution stress test on Piscis [18] smFISH data showed that PhySeqNet can generalize zero-shot to images with similar optical and emitter characteristics, but performance degrades when background and noise statistics differ substantially from the training distribution. Users applying PhySeqNet to new microscope platforms should retrain with noise frames acquired on their specific instrument.

Future directions include integration of a neural basecaller that can learn sequence-dependent effects such as phasing, bleedthrough, and signal decay across sequencing cycles to reduce preprocessing overhead in pyBARseq.

## Methods

### 1. Sample preparation, imaging, and the pyBARseq pipeline

BARseq was performed on a coronal section from a C57BL/6J, male, ~4 months old mouse brain. Brain was flash-frozen, cryosectioned at 20 μm, and processed through the BARseq chemistry [8]. Imaging was performed on an inverted microscope (Nikon Ti2) with a spinning disk module (CrestOptics V3) attached to the output port. A 20x/0.7NA objective was employed for the acquisition and a Kinetix (Teledyne Photometrics) camera was used, yielding a lateral pixel size of 0.33 μm and an axial step of 1 μm. The section was tiled across 60 fields of view (FOVs) of 3200 × 3200 pixels with 23% overlap between adjacent tiles. Maximum-intensity projections (MIPs) over the imaged z-stack (30 μm, 1μm step-size) were computed for each FOV and channel and used as input to all subsequent processing. Gene-panel and imaging channel details are mentioned in Supplementary note 1.

pyBARseq is a modular Python package that takes raw multi-cycle, multi-channel sequencing tiles through to a cell-by-gene expression matrix and an interactive napari QC viewer. The pipeline operates per-tile through preprocessing (Noise2Void denoising [35], morphological background subtraction, calibrated channel alignment, linear spectral bleedthrough correction, and block-correlation cycle registration), then aligns hybridization and barcode-sequencing cycles to the geneseq reference, stitches tiles using FIJI/MIST [36] and segments cells with Cellpose v3 (cyto3, default diameter 40 px) [37]. Spot calling is performed by BarDensr [16] or through PhySeqNet as per user input; hybridization cycles are decoded by per-channel h-maxima detection. The accompanying Napari [38] plugin pyBARseq Viewer exposes every pipeline intermediate for interactive QC and gene exploration.

### 2. Physics-informed generation of semi-synthetic rolony images

Training data for the fluorescent foci detector were generated by forward simulation of a physical imaging model and the addition of real, tissue-free experimental noise, yielding semi-synthetic images for which exact emitter coordinates are known.

PSF based on Born and Wolf model [26, 33] applied to a confocal system (discussed in Results Section) was implemented in MATLAB. To capture the optical variability present across acquisitions in similar systems, a bank of 210 channel-specific PSFs was precomputed by jittering two parameters of the PSF. The numerical aperture was sampled on a uniform grid of 21 values from 0.5 to 0.7 in steps of 0.01, upto the nominal NA of the imaging objective; for each NA value, the effective pinhole diameter was independently jittered ten times by drawing from a uniform distribution on [40, 60] μm centered on the physical pinhole size of 50 μm. Wavelengths ($\lambda_exc$, $\lambda_em$) were held fixed at their channel-specific values (520/550, 540/590, 640/690, 640/730 nm for G, T, A, C detection channels). Because the symmetric PSF model could not fully account for a small channel-dependent asymmetry observed in the averaged profiles (most pronounced in the red channels A and C in our system), channel-specific PSF anisotropy was estimated from averaged 1μm fluorescent bead images acquired in each sequencing channel by fitting each bead profile with a rotated 2D Gaussian (Supplementary note 2). The resulting channel-specific stretch parameters were then used to approximate the dominant aberration distortion in the simulation model. This procedure estimates an effective PSF anisotropy relevant to the observed bead images, rather than a full physical aberration model. At training time, a single PSF was drawn uniformly at random from the channel-appropriate bank for each simulated image, so that NA and pinhole variability appeared to the network as inter-image rather than intra-image variation. ( Supplementary fig. S1). Although the simulator supports separate PSF generation for all four sequencing channels, the detector training used two wavelength-band-specific models rather than four independent models. Model G was trained using the G-channel simulation parameters and applied to both G and T images, while Model C was trained using the C-channel simulation parameters and applied to both A and C images.

For each simulated training image, a high-resolution three-dimensional fluorescent emitter volume was generated at isotropic sampling $\delta x = \delta z = 0.1 \mu m$. The axial simulation range spanned −20 to +20μm. This axial support was intentionally broader than the 20 μm section used in the present experiment, allowing the simulator to represent a wider range of emitter depths and section-thickness conditions. Effective emitter diameters were sampled from a log-normal distribution clamped to the range $[9,25]$ high-resolution voxels. This corresponds to effective physical emitter diameters of 0.9-2.5 μm. The term "effective diameter" denotes the support/scale of the simulated fluorescent source distribution rather than a hard physical rolony diameter. One hundred sizes were pre-sampled at the start of generation and indexed randomly per emitter. Each emitter was modeled as a 3D Gaussian truncated at radius $r_d$, with intensity set to zero outside this support and the resulting profile normalized to unit peak intensity.

$$\rho(\mathbf{r}) = \exp\left(-\frac{|\mathbf{r}|^2}{2(\sigma_d)^2}\right) \cdot \mathbb{1}[|\mathbf{r}| \leq r_d],$$

where $r_d = round(d/2) - 1$ voxels and $\sigma_d = max(1, r_d/2)$.

Two populations were generated per image. The first comprised 100 isolated emitters placed at random positions (representative of extra-somal rolonies) within the volume. The second comprised 2-20 dense "cell-like" clusters, each containing 50-100 emitters placed at random offsets of up to ±50 high-resolution voxels along each spatial axis from a cell anchor point. These numbers were chosen empirically to approximate the density and spatial organization observed in real BARseq images. To avoid unrealistically collapsed source centers, candidate emitter positions were resampled up to 500 times when their center-to-center distance was less than approximately half the sum of their effective diameters. This procedure suppressed severe source overlap while still allowing dense packing of soft, deformable fluorescent source distributions. Each rolony was assigned an intensity proportional to $\eta m$, where $\eta = 100$ and $m$ was sampled uniformly from integers between $n_{\text{molecule}} - 5$ and $n_{\text{molecule}} + 5$, with $n_{\text{molecule}} = 15$. This yielded per-rolony brightness multipliers between 1000 and 2000.

The high-resolution emitter volume was convolved with the calibrated, aberration-mimicked per-channel PSF using FFT-based 3D convolution, with the PSF normalized to unit sum. The convolved volume was downsampled to the camera grid by 3D box-kernel interpolation at downsampling ratios $\delta x/0.33 = 0.303$ laterally and $\delta z/1.0 = 0.1$ axially, followed by maximum-intensity projection along $z$ to produce a 256 × 256 simulated MIP.

A random crop of size 256 × 256 was sampled from the empty-frame pool and added pixel-wise to the simulated MIP.

The corresponding target map was generated from the simulated emitter center coordinates. For each emitter, the high-resolution voxel at its centroid location was incremented by one. This 3D centroid map was downsampled to the camera grid using the same spatial scale factors and sum-projected along z to obtain a 2D target map. The projected target was binarized before training so that each positive pixel indicated the presence of one or more simulated emitter centers. In addition, a density map was also prepared before training to emphasize dense emitters (Supplementary note 3).

### 3. Training a physics-informed U-Net for sparse emitter source detection

Model G was trained on 3876 images, validated on 620 images and tested on 320 images. Model C was trained on 5269 images, validated on 486 images and tested on 551 images. These totals include both simulated emitter-containing images and empty-frame-only images ($I_{\mathrm{sim}} = 0$), which were included to provide background-only examples and reduce false-positive detections in sample-free regions.

The detector was a 2D U-Net [39], implemented in PyTorch 2.4.1 with CUDA 12.1 build, with single-channel grayscale input of size 256 × 256, depth $D = 4$ (four down/upsampling stages), and base channel count $C_0 = 64$ (Supplementary fig. S2). Two independent models were trained, one per channel pair: Model G covered channels G and T (green-yellow region) and Model C covered channels A and C (red region). The two models shared the architecture above with one difference: Model C included spatial dropout (Dropout2d) inserted after the LeakyReLU activations in the deepest two encoder levels (rate 0.1) and after the second convolution of the bottleneck (rate 0.2). Model G was trained without dropout (Supplementary note 4). Inference is performed on 256 x 256 cropped overlapping patches and stitched back to obtain full tile detections (Supplementary note 6).

The training loss combined a density-weighted binary cross-entropy with a background-sparsity regularizer. The dominant term was a density-weighted BCE-with-logits, where two complementary weighting schemes acted together: a global positive-class weight $w_{\mathrm{pos}}$ addressed the class imbalance and a spatial weight map $W(x, y)$ derived from the local rolony density (Supplementary note 3) up-weighted pixels in dense clusters where detection is most difficult. The $\mathcal{L}_{\mathrm{bg}}$ term reinforces background suppression specifically on negative pixels. The full loss was

$$\mathcal{L} = \underbrace{\frac{\sum_i W_i \cdot \mathrm{BCE}_{w_{\mathrm{pos}}}(z_i, t_i)}{\sum_i W_i + \varepsilon}}_{\mathcal{L}_{\mathrm{det}}} + \underbrace{\lambda_{\mathrm{bg}} \cdot \frac{1}{|\mathcal{B}|} \sum_{i \in \mathcal{B}} \sigma(z_i)}_{\mathcal{L}_{\mathrm{bg}}},$$

where $z_i$ is the predicted logit at pixel $i$, $t_i \in \{0,1\}$ the binarized target, $\sigma$ the sigmoid function, $\mathcal{B} = \{i : t_i = 0\}$ the set of background pixels, $\mathrm{BCE}_{w_{\mathrm{pos}}}$ binary cross-entropy with logits and positive-class weight $w_{\mathrm{pos}}$, and $\varepsilon = 10^{-8}$ stabilizes the normalization. Both weighting parameters were selected by Optuna hyperparameter search [40] (Supplementary note 5): $w_{\mathrm{pos}} = 1.7$ and $\lambda_{\mathrm{bg}} = 0.126$.

Inputs were min-max normalized per image to $[0,1]$ in the dataloader,

$$\hat{I} = \frac{I - \min(I)}{\max(I) - \min(I) + \varepsilon},$$

with $\varepsilon = 10^{-10}$, before center-cropping to 256 × 256. Models were trained with the Adam optimizer [41] at learning rate $4 \times 10^{-4}$, $\beta_1 = 0.912$, $\beta_2 = 0.92$, and weight decay $6.5 \times 10^{-6}$, with a StepLR schedule decaying the learning rate by a factor of 0.77 every 14 epochs. Training ran for 100 epochs at batch size 4 on a single NVIDIA RTX A6000 GPU. No data augmentation was applied beyond the natural variation already present in the simulation. Final models were selected at epochs after which validation and train loss showed no improvement in convergence.

**4. PhySeqNet: sparse U-Net detection coupled to argmax barcode decoding**

PhySeqNet refers to the combination of the U-Net detector (Section 3) with a max-channel argmax basecaller. Detection and basecalling are decoupled: detection is performed once per channel for the first cycle, while basecalling operates over the resulting sparse coordinate set across all 7 cycles.

For each detection coordinate in the union of per-channel binary detection maps, a 4 × 7 intensity vector was extracted by sampling the preprocessed image stack at that coordinate across all 4 channels and 7 cycles. Per-cycle base calls were obtained by argmax over the 4 channels:

$$\hat{c}(t) = \arg\max_{c \in \{1,2,3,4\}} I_c(t), \quad t = 1, \dots, 7,$$

producing a called 4-bit-per-cycle barcode $\hat{\mathbf{T}}$. The XOR Hamming distance to every codebook entry was computed,

$$d_g = \sum_{t=1}^{7} \sum_{c=1}^{4} \mathbf{T}_g\,(c,t) \oplus \hat{\mathbf{T}}(c,t),$$

and the gene assignment was taken as the codebook entry with minimum distance, $\hat{g} = \mathrm{argmin}_g d_g$, with the corresponding error $d_{\hat{g}}$ retained for downstream filtering.

Since the PhySeqNet detections are performed on raw images without spectral bleedthrough correction, the 4 channels combined produce ghost detections for bleedthrough cases. However, the basecalling is performed on bleedthrough corrected images which enables easy detection and removal of these ghost detections. Two channel labels were computed per detection coordinate: a pre-detection label encoded as a power-of-2 over channels in which the cycle-1 detection map fired (1, 2, 4, 8 for single-channel detections; non-power-of-2 values for multi-channel detections), and a post-call label given by the called channel at cycle 1. All channel detections are summed into a single channel image to select all rolony coordinates irrespective of their channels. A coordinate was retained if and only if either:

- the detection was uniquely in one channel and the pre- and post-call labels matched ($\mathrm{count} = 1$, $\mathrm{pre_{ch}} = \mathrm{post_{ch}}$), or
- the detection was simultaneously in multiple channels ($\mathrm{count} = 2$) because the final single-channel detection mask will retain the dominant channel,

both subject to the error threshold $d_{\mathrm{XOR}} < 6$ (Supplementary note 7):

$$[d_{\mathrm{XOR}} < 6 \wedge \mathrm{count} = 1 \wedge \mathrm{pre_{ch}} = \mathrm{post_{ch}}] \vee [d_{\mathrm{XOR}} < 6 \wedge \mathrm{count} = 2].$$

Surviving coordinates are the raw PhySeqNet output.

After barcode assignment, detections were subjected to barcode-aware greedy non-maximum suppression (NMSB) to remove closely spaced duplicate candidates without merging distinct barcode rolonies. Detections were grouped by decoded barcode and ranked by their associated PhySeqNet detection score. Beginning with the highest-scoring detection, lower-scoring detections assigned to the same barcode with a Euclidean center-to-center distance <2 pixels (<0.66 μm) were suppressed. Detections assigned to different barcodes were not mutually suppressed. This procedure was repeated until no suppressible same-barcode detections remained, and the retained coordinates constituted the PhySeqNet_NMSB output.

**5. Real BARseq called-spot benchmark**

Real BARseq detection outputs were benchmarked on 10 edge-trimmed fields of view to ensure a common evaluation domain across methods. PhySeqNet (with and without NMSB) and BarDensr detections were loaded from their corresponding called-spot CSV files, while LoG, DoG, and h-maxima baselines were applied directly to the same image. Baseline detector thresholds were tuned separately for each field of view by maximizing forward-model normalized cross-correlation on a central 800 × 800 pixel crop and then applied to the full edge-trimmed image. Piscis was additionally evaluated as a generic pretrained spot-proposal baseline using the 20251212 model with threshold 0.5 [18]. BarDensr was operated using the empirically-determined optimal noise threshold for our system [16]. Cross-method overlap was quantified using a 2-pixel colocalization radius. To assess physical consistency, each method's detections were converted to a sparse point map weighted by the underlying image intensity, convolved with a Gaussian PSF kernel with σ = 1.5 pixels, and compared with the original image using normalized cross-correlation (NCC) computed per tile as.

$$\mathrm{NCC}(\boldsymbol{I}, \hat{\boldsymbol{I}}) = \frac{\sum_p (I_p - \bar{I})(\hat{I}_p - \bar{\hat{I}})}{\sqrt{\sum_p (I_p - \bar{I})^2} \cdot \sqrt{\sum_p (\hat{I}_p - \bar{\hat{I}})^2}}$$

where $I_p$ is the raw image intensity at pixel $p$, $\hat{I}_p$ is the forward-model reconstruction at pixel $p$, $\bar{I}$ and $\bar{\hat{I}}$ are their pixel means.

NCC and MS-SSIM were summarized across the 10 matched fields of view, with pairwise comparisons performed using one-sided paired Wilcoxon signed-rank tests followed by Benjamini-Hochberg correction for six comparisons within each metric.

Multi-scale structural similarity (MS-SSIM) was computed between each original image and its corresponding forward-model reconstruction using identical preprocessing and intensity normalization across methods. MS-SSIM evaluates luminance, contrast, and structural agreement over progressively downsampled image scales, with larger values indicating greater structural similarity [34]. The same implementation and parameter settings were used for all methods. For the external zero-shot benchmark, local Piscis datasets were organized into visually comparable punctate/RNA-FISH-like datasets and excluded datasets with mismatched morphology or channel content [18]. Raw images, ground-truth point annotations, and PhySeqNet binary prediction masks were loaded from each selected dataset folder. The primary metric was ground-truth-centered coverage: a ground-truth spot was counted as recovered if at least one positive mask pixel occurred within a 3-pixel radius. Piscis predictions were generated on the same raw images using the pretrained 20251212 model with threshold 0.5 and deduplicated with a 1.5-pixel radius. Precision, recall, and F1 were computed at a 3-pixel matching radius.

### 6. Reference-guided cell-type annotation using scANVI

Reference Allen Brain Cell Atlas dataset for mouse brain [30] (10x v3 single cell RNA-seq dataset, Supplementary note 9) and BARseq query cells were integrated using scVI (scvi-tools 0.16.3) [28], batch labels from the dataset/modality annotation, and per-cell size factors computed from batch-normalized library sizes. scVI/scANVI models were configured with a 20-dimensional latent space, two fully connected hidden layers, 128 hidden units per layer, and a negative binomial gene-expression likelihood for 200 epochs with early stopping on validation ELBO [28]. scANVI was initialized from the trained scVI model using the same registered AnnData setup and trained for an additional 100 epochs separately for each label hierarchy. For each query cell, the predicted scANVI label, full class-probability vector, prediction entropy, and maximum class probability were retained. A confidence threshold of 0.7 was used to summarize high-confidence predictions and generate confidence-stratified cell-type count comparisons. Companion MERFISH data were loaded from a matched coronal MERFISH section from the Allen Brain Cell Atlas for mouse brain [30, 31]. BARseq-MERFISH comparisons were restricted to the shared gene set after excluding the three hybridization markers.

### Data and code availability

pyBARseq, napari plugin, trained U-Net models (Models G and C), the simulation MATLAB code and all analysis code reproducing the figures in this paper will be made available upon publication at https://github.com/NRG-DDN/PyBARseq. Raw and processed BARseq data will be made available upon publication. The Allen MERFISH reference data are available through the Allen Brain Cell Atlas portal (https://portal.brain-map.org/).

## Declarations

*Funding*

This work was supported by the National Institutes of Health under awards RF1MH126883 and U01NS132353 (to A.M.Z.) and U19NS123716 (to A.M.Z.).

*Author Contributions*

N.G. conceived the study, developed PhySeqNet, built the pyBARseq Python pipeline, performed all computational analyses, created all figures, and wrote the manuscript. R.G. performed BARseq library preparation, sequencing chemistry, and image acquisition. Y.W. performed brain extraction, cryosectioning, and animal handling. H.Z. supervised BARseq core operations and quality control. X.C. provided guidance on BARseq pipeline design and analysis. J.H. contributed to pyBARseq software development. A.M.Z. supervised the study and edited the manuscript.

## Supplementary Note 1: Codebook and imaging channels

The section was probed with a panel of 109 genes comprising 106 sequencing genes and 3 hybridization (hyb) genes (Slc17a7, Gad1, Slc30a3). Sequencing-by-synthesis was carried out for 7 cycles using the four-channel chemistry (G, T, A, C), followed by a single hybridization round for the three hyb markers with an additional hyb probe included for all genes and DAPI for nuclear staining [1].

## Supplementary Note 2: Beads calibration

For each channel, beads were segmented from the MIP by morphological opening, locally adaptive thresholding, and a size filter on EquivDiameter, yielding (~1,700–1,900) beads per channel. A 31 × 31 px crop was extracted around each bead centroid and averaged to produce a per-channel mean bead profile. The fitted major and minor Gaussian widths provided the effective stretch ratio of the PSF, and the fitted rotation angle provided the direction of elongation. Because the fitted rotation angles were small for all four channels the correction was implemented in subsequent training-data generation as an axes-aligned x-stretch only, with sx = 1.00, 1.00, 1.12, 1.16 for G, T, A, C (sy = 1). The per-channel distortion stretch ratios extracted from real bead fits were applied deterministically to each PSF rather than jittered, since these reflect calibrated optical-path properties rather than session-to-session variability. Other potential sources of PSF variability such as refractive-index mismatch, off-axis distortion, higher-order aberrations, and depth-dependent spherical aberration were not modeled, on the rationale that the dominant residual mismatch between the symmetric Born model and real bead profiles was already captured by the per-channel shape distortion correction. The full four-channel PSF calibration was retained to define the optical model and to support future channel-specific training if needed.

## Supplementary Note 3: Density weight map

In addition to the binary detection target, a density-derived spatial weight map was generated and used during training as a per-pixel weight on the detection loss. For each target image, foreground pixels were identified by thresholding the projected target map at >0, and connected components were extracted. For each connected component, the number of foreground pixels within a 10-pixel square neighborhood centered on the component centroid was counted. Pixels belonging to that component were assigned a weight based on this local foreground density, while background pixels retained weight 1. The resulting weight map was clipped to the range {1,2,3} such that isolated or low-density detections received weight 2 and locally crowded detections received weight 3.

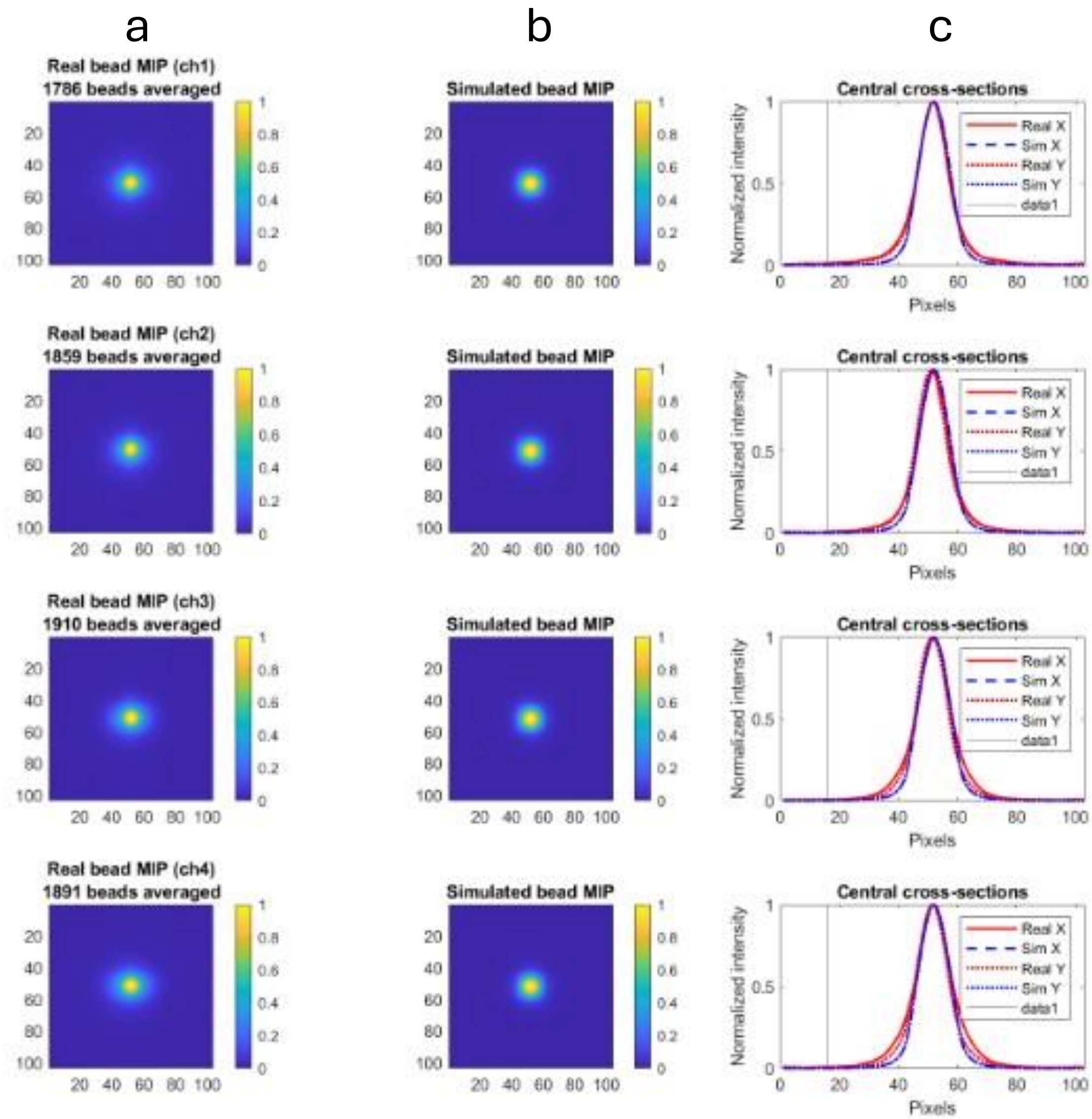

Figure S1. Calibration of the simulated point-spread function using experimental fluorescent beads. Rows show the G, T, A, and C fluorescence channels. From left to right, panels show the mean experimentally measured bead profile, the corresponding symmetric simulated bead profile generated from the optical model, and central horizontal and vertical intensity cross-sections. Channel-dependent PSF anisotropy measured from the bead data was incorporated into the simulation as an axis-aligned stretch, with the strongest corrections required for the A and C channels.

### Supplementary Note 4: U-Net architecture

The 2D U-Net [2] operates on single-channel grayscale input of size 256 × 256 and has depth 4 and base channel count 64 (doubling at each encoder level (encoder channels 1 → 64 → 128 → 256 → 512, bottleneck 1024, decoder 512 → 256 → 128 → 64, final 1 × 1 convolution to a single-channel logit map). Each encoder and decoder unit consisted of two convolutional blocks, each comprising a 3 × 3 Conv2d with padding='same', an InstanceNorm2d layer with affine parameters, and a LeakyReLU activation with negative slope 0.2. Downsampling between encoder levels used MaxPool2d(2, 2); upsampling in the decoder used bilinear Upsample followed by a 3 × 3 convolution. Skip connections concatenated encoder feature maps to the corresponding decoder inputs before the decoder convolutions. Convolution weights were initialized by Kaiming normal initialization (nonlinearity='leaky_relu', a=0.2) and biases by zero; normalization layers were initialized with weight 1 and bias 0

**Supplementary Note 5: Hyperparameter search**

A hyperparameter search for the base model applicable to both model G and model C was performed using Optuna 4.5.0 [3] with a Tree-structured Parzen Estimator (TPE) sampling and median pruning, optimizing validation binary F1 score across 100 trials. The search space included positive-class weight $w_{pos}$, background penalty $\lambda_{bg}$, learning rate, Adam $\beta_1$and $\beta_2$, weight decay, learning-rate scheduler type and parameters, U-Net depth, base channel number and detection threshold. The selected best-trial values were $w_{\mathrm{pos}} = 1.7$, $\lambda_{\mathrm{bg}} = 0.126$, learning rate $4 \times 10^{-4}$, $\beta_1 = 0.912$, $\beta_2 = 0.92$, weight decay $6.5 \times 10^{-6}$, step period 14 epochs, decay $\gamma = 0.77$, depth 4, base channels 64, and detection threshold 0.733. For Model C, encoder dropout rate 0.1 (in the deepest two encoder levels) and bottleneck dropout rate 0.2 were added empirically.

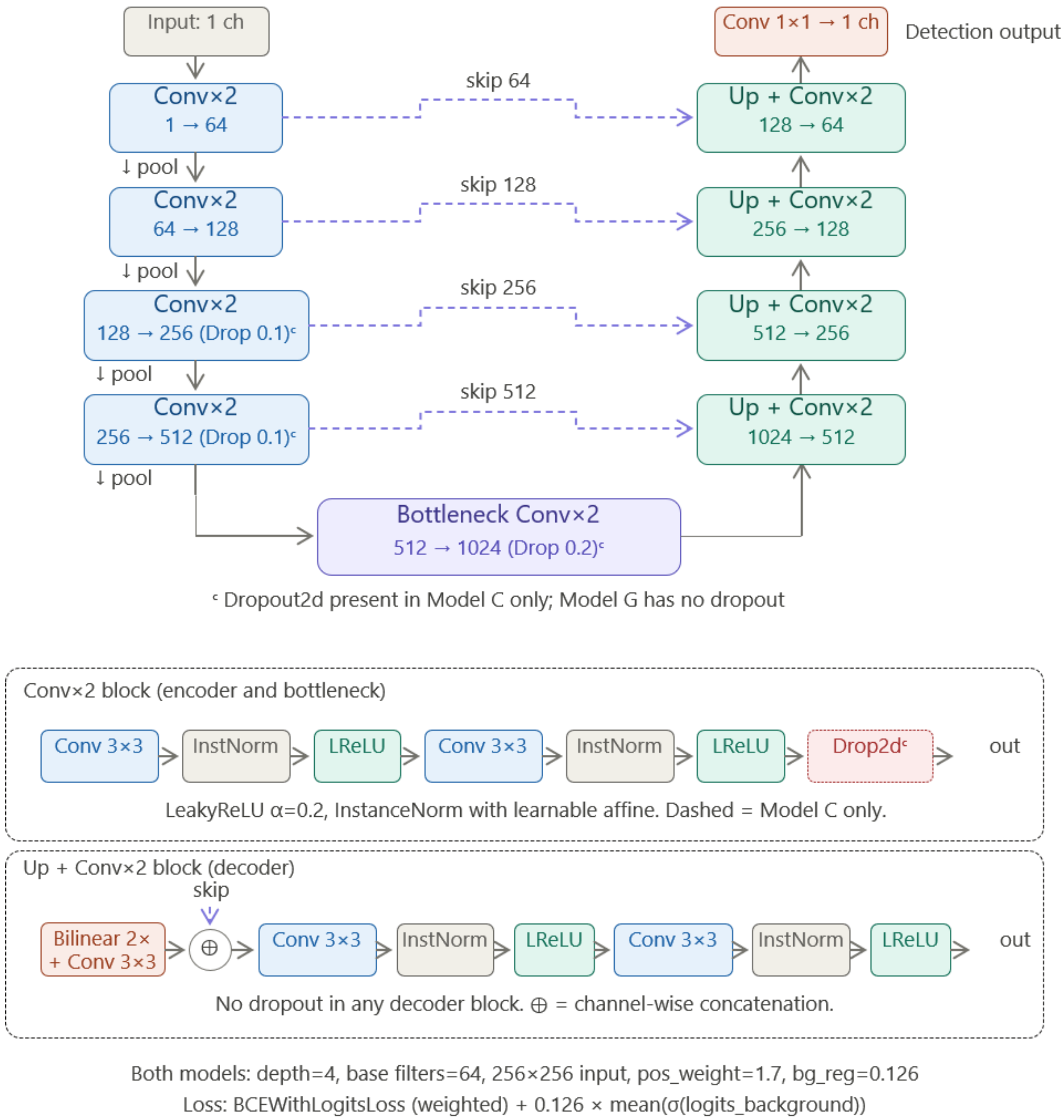

Figure S2. U-Net architecture used for PhySeqNet emitter detection. The network accepts a single-channel 256 × 256 image and uses a four-level encoder-decoder architecture with skip connections and a single-channel logit output. Each convolutional block consists of 3 × 3 convolutions, InstanceNorm, and LeakyReLU activations; decoder upsampling is performed by bilinear interpolation followed by convolution. Model C additionally includes Dropout2d in the two deepest encoder levels and bottleneck, whereas Model G is trained without dropout. The lower schematics show the encoder/bottleneck and decoder block structures.

## Supplementary Note 6: Inference on real data

At inference time, full FOVs (3200 × 3200 px) were processed in overlapping patches. Each FOV was tiled into 256 × 256 patches with stride 224 (32-px overlap), with an additional flush patch added at each row and column boundary if the regular grid did not reach the FOV edge. Each patch was min-max normalized identically to training, passed through the appropriate model (G or C), and the resulting sigmoid probability map was thresholded at 0.733 to produce a binary detection map per patch. Detections were stitched back to FOV coordinates. For overlapping strips in inference patches, detections near tile boundaries were assigned to the patch whose center was closest to the detected pixel, thereby avoiding duplicate detections from neighboring patches. The stitched binary detection map was retained as a per-FOV 4-channel image (one channel per sequencing channel), and used as input to the basecaller.

## Supplementary Note 7: Codebook representation and the bit-error/2 convention for PhySeqNet

The codebook provided 114 sequence x gene assignments (106 genes + 5 unused barcodes+ 3 hybridization genes), each a 7-mer over the alphabet $\{G, T, A, C\}$ except the hybridization genes which carry channel numbers as the code. For algorithmic processing, each base was encoded as a 4-bit one-hot vector:

$$G \mapsto [0,0,0,1], \quad T \mapsto [0,0,1,0], \quad A \mapsto [0,1,0,0], \quad C \mapsto [1,0,0,0],$$

so that each barcode became a 4 × 7 binary matrix $\mathbf{T}_g \in \{0,1\}^{4\times 7}$ for gene $g$.

Under this encoding, two distinct bases differ in exactly two bits (one bit flipped from 0 to 1, one from 1 to 0). Consequently, the bitwise XOR Hamming distance between two encoded barcodes equals **twice** the number of base-position substitutions:

$$d_{\mathrm{XOR}}(\mathbf{T}_a, \mathbf{T}_b) = 2 \cdot d_{\mathrm{base}}(\mathbf{T}_a, \mathbf{T}_b).$$

All thresholds in this paper are stated as XOR Hamming distances to match the values used in code. A threshold of $d_{\mathrm{XOR}} < 6$ thus corresponds to fewer than 3 base-position substitutions.

## Supplementary Note 8: FDR estimation via unused barcodes

The codebook contained 5 barcode sequences with no assigned gene (“unused” codes). Detections assigned to unused barcodes are by construction false positives. The empirical FDR was estimated as [1]

$$\widehat{\mathrm{FDR}} = \frac{N_{\mathrm{unused_calls}}/N_{\mathrm{unused_codes}}}{N_{\mathrm{all_calls}}/N_{\mathrm{used_codes}}},$$

where $N_{\text{unused_calls}}$ and $N_{\text{all_calls}}$ are the total detections assigned to unused and all barcodes respectively and $N_{\text{unused_codes}}$ and $N_{\text{used_codes}}$ are the total unused and used barcodes in the codebook respectively.

**Supplementary Note 9: Reference Allen dataset**

The reference was the Allen Brain Cell Atlas 10x Chromium v3 single-cell RNA-seq dataset of mouse brain [4]. The reference was provided in two label hierarchies: class and subclass with supertype excluded because of the limited gene panel in BARseq. All downstream analyses were restricted to the BARseq panel genes.

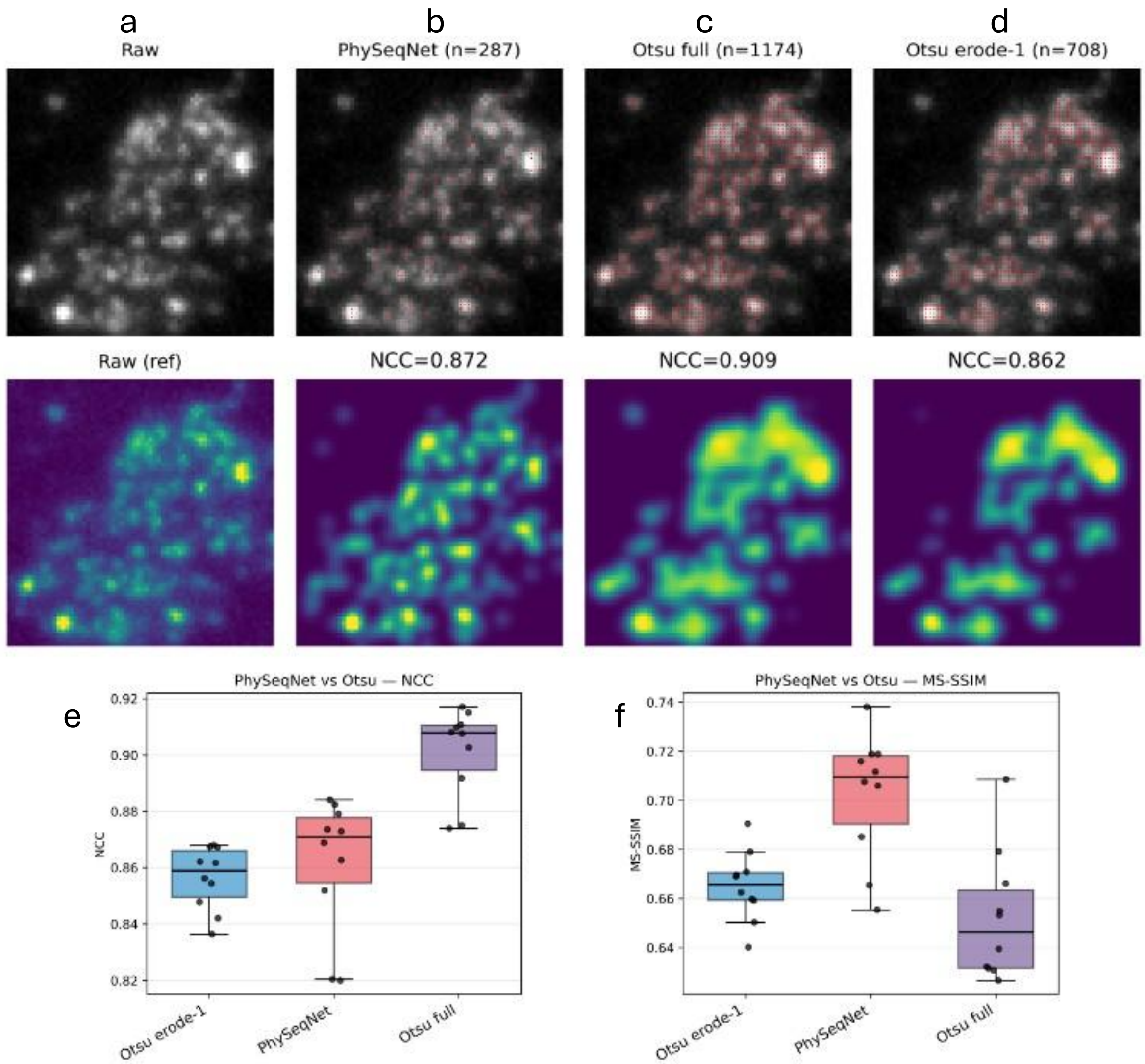

Figure S3. Forward-model reconstruction distinguishes sparse emitter detection from foreground segmentation. a–d, Representative dense BARseq field comparing the raw image with PhySeqNet detections, full Otsu foreground thresholding, and Otsu thresholding followed by one-pixel erosion. The corresponding forward-model

reconstructions are shown below each detection map. Full Otsu thresholding produces substantially more positive coordinates and achieves a high NCC by representing extended foreground regions, whereas PhySeqNet produces a sparse set of discrete emitter coordinates. e,f, NCC and MS-SSIM across evaluated fields for PhySeqNet and the two Otsu baselines. Although full Otsu thresholding can achieve higher NCC, PhySeqNet yields higher MS-SSIM and a substantially sparser source representation.

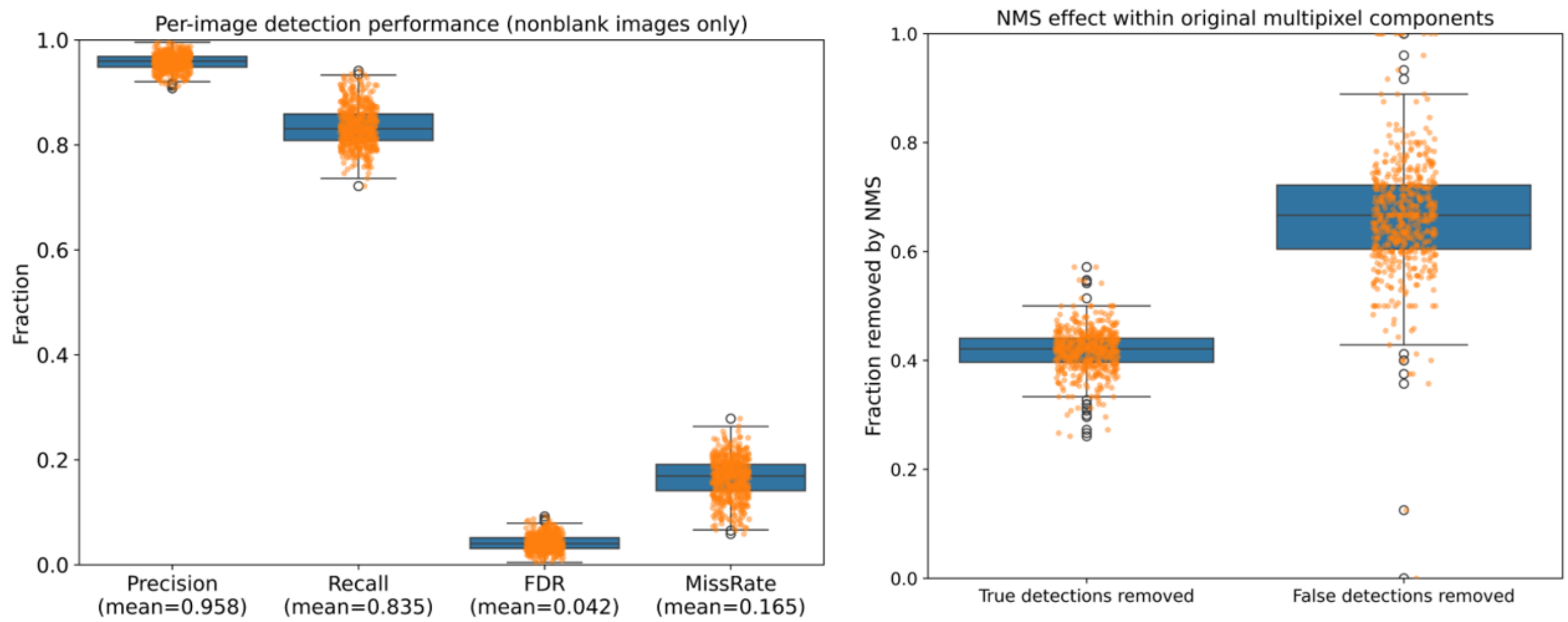

Figure S4. NMSB preferentially suppresses false detections while reducing detection sensitivity. a, Per-image detection performance after NMSB on nonblank held-out simulated images, showing mean precision of 0.958, recall of 0.835, false-discovery rate of 0.042, and miss rate of 0.165. b, Fraction of simulated true and false detections removed by non-maxima suppression within originally multipixel detection components. NMS removes a larger fraction of false detections than true detections, illustrating the precision-sensitivity tradeoff introduced by the conservative suppression step.

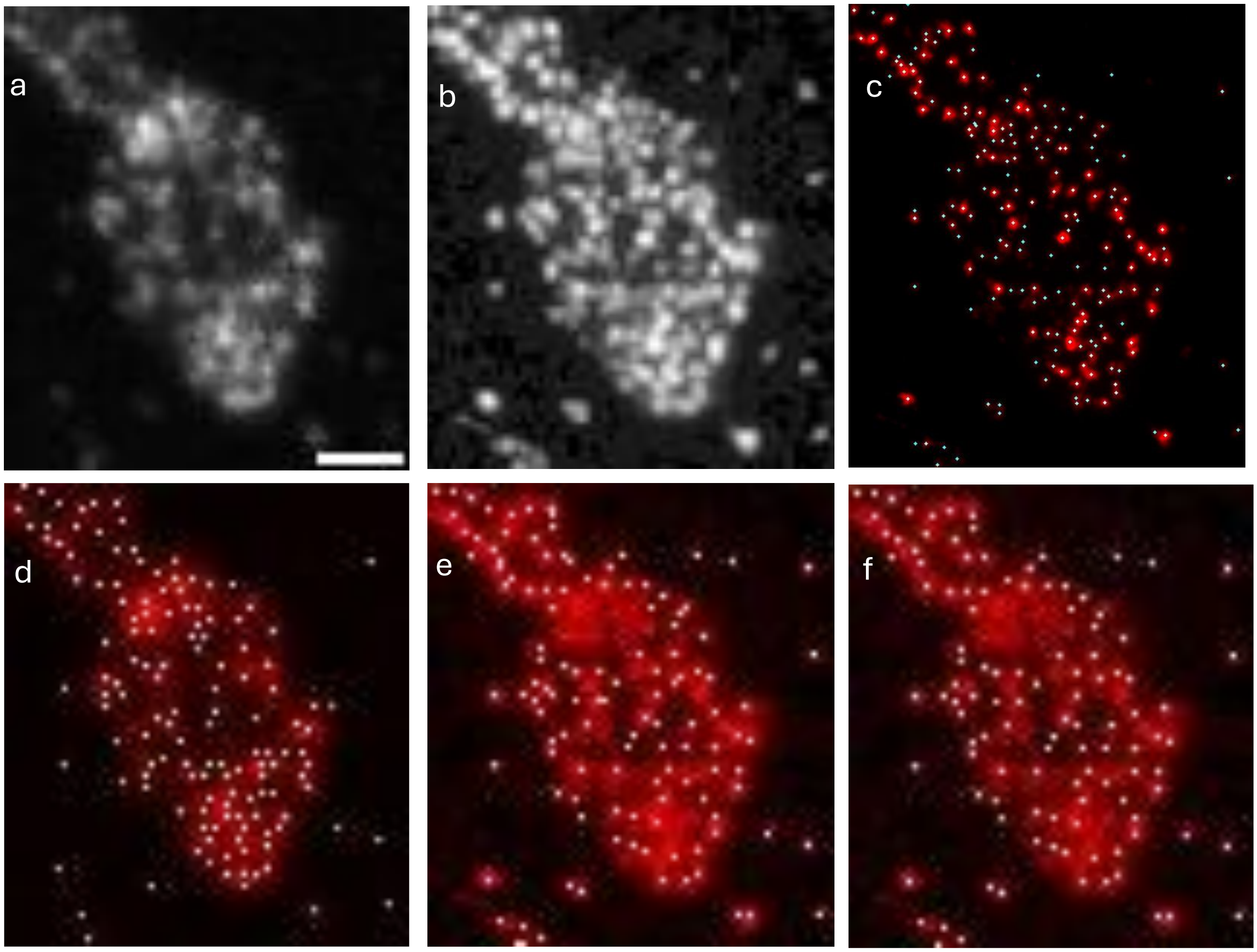

Figure S5. High-resolution imaging provides an independent validation of PhySeqNet source localization. a, Representative region from the original 20x/0.7NA BARseq acquisition. b, The corresponding region imaged at higher resolution using a 63x/1.4NA Airyscan objective, registered and resampled to the lower-resolution coordinate system. c, Independent LoG detections obtained from the raw 63x/1.4NA image. d, PhySeqNet detections after NMSB on the original 20x/0.7NA BARseq image. e,f, PhySeqNet detections on the resampled high-resolution image before and after NMSB, respectively. The spatial correspondence between PhySeqNet detections in the original BARseq image and puncta resolved by the higher-resolution acquisition provides an orthogonal validation of source localization, while application to the high-resolution image illustrates transfer outside the acquisition conditions used for training. Scalebar in a, 5 µm.

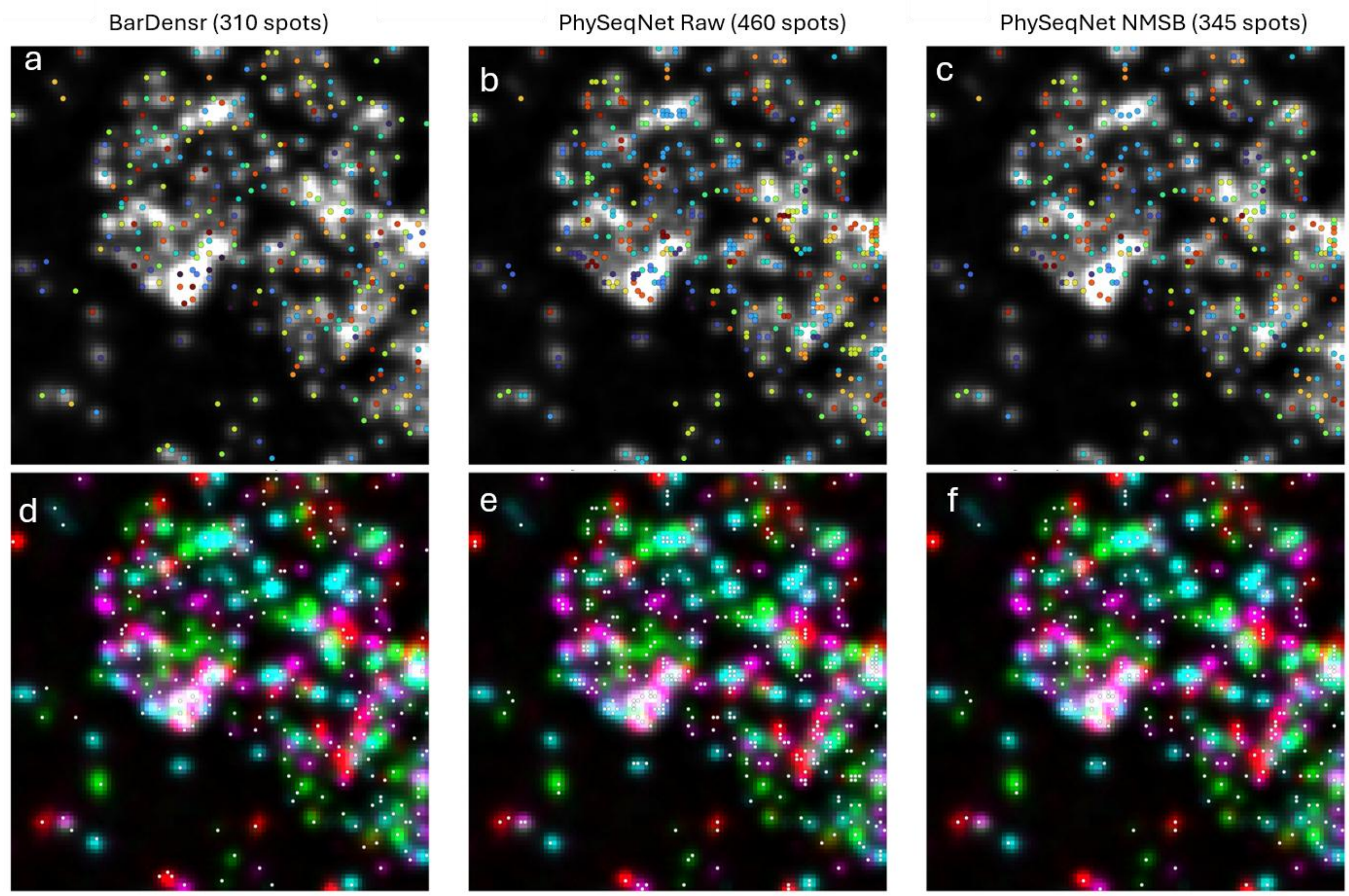

Figure S6. NMSB reduces dense PhySeqNet predictions while preserving the underlying detection pattern. Representative dense BARseq region comparing  single-shot density estimation from BarDensr, raw PhySeqNet, and PhySeqNet after NMSB. a-c, Detection coordinates overlaid on a representative fluorescence image, colored by gene identity; the corresponding spot counts are 310, 460, and 345, respectively. d-f, The same detection coordinates shown on the multichannel fluorescence image with each pseudocolor representing one nucleotide channel (G, T, A, C). Raw PhySeqNet produces a denser candidate set than BarDensr, whereas NMSB collapses closely spaced same-barcode candidates to produce a more conservative final detection set.

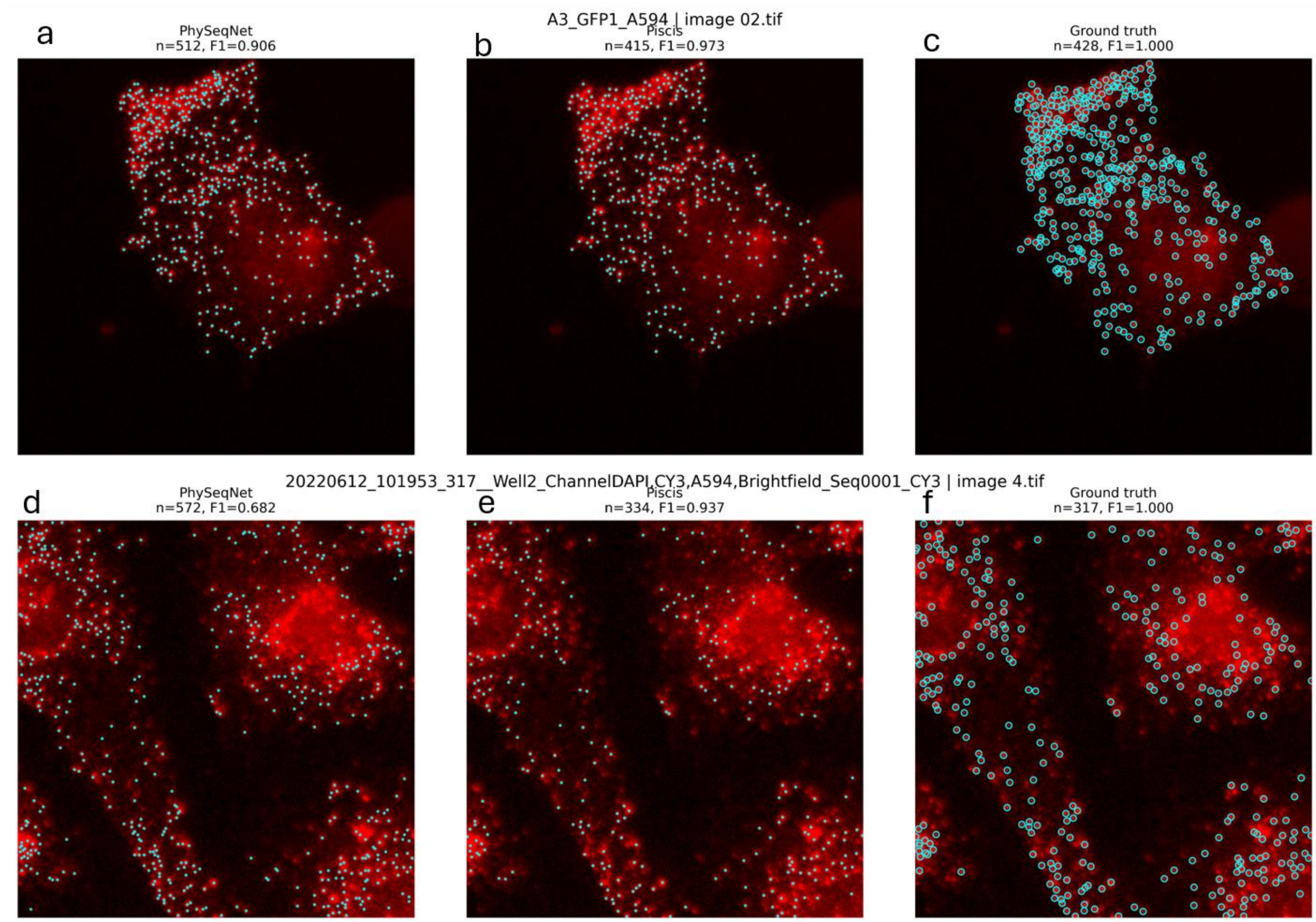

Figure S7. Zero-shot PhySeqNet detection on out-of-distribution Piscis RNA-FISH benchmark images [5]. Two representative benchmark images with punctate fluorescence and comparatively favorable signal-to-background conditions are shown. a,d, PhySeqNet predictions obtained without retraining on the Piscis datasets. b,e, Predictions from Piscis, evaluated in its native domain. c,f, Reference ground-truth annotations. PhySeqNet achieves F1 scores of 0.906 and 0.682 on the two images, respectively, despite being trained exclusively on BARseq-oriented simulations, indicating partial transfer of the learned source-localization behavior across imaging modalities.

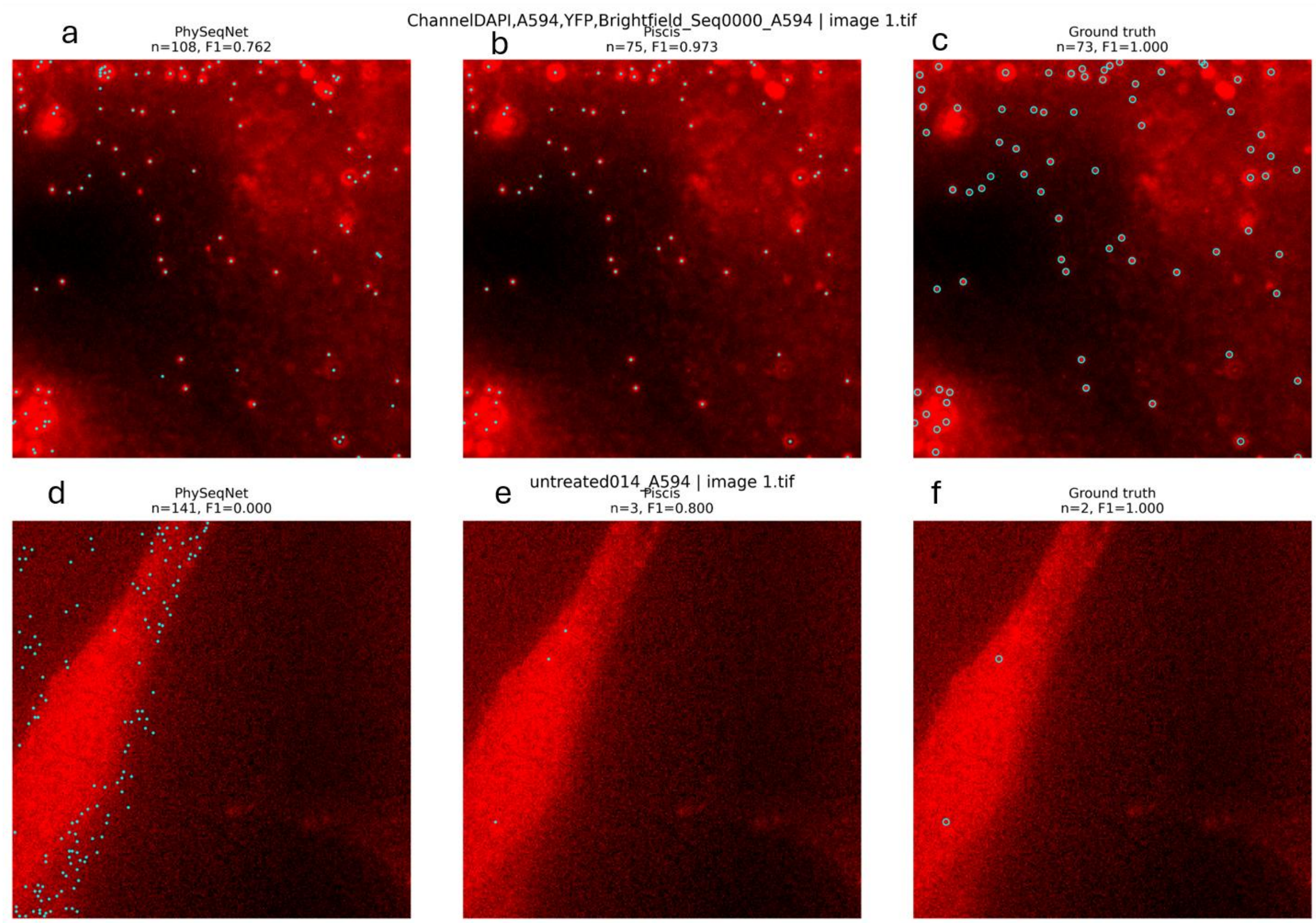

Figure S8. Out-of-distribution performance degrades when background statistics differ substantially from the PhySeqNet training domain. a-c, out-of-distribution Piscis RNA-FISH benchmark images [5] with increased diffuse background, showing PhySeqNet predictions, Piscis predictions, and reference annotations, respectively. PhySeqNet retains partial detection performance in this condition. d-f, Near-empty field dominated by structured diffuse background, where the BARseq-trained PhySeqNet model produces numerous false detections and fails to reproduce the sparse reference annotation, whereas Piscis remains effective in its native data domain. These examples illustrate the importance of matching the simulated training noise and background distribution to the target imaging modality.

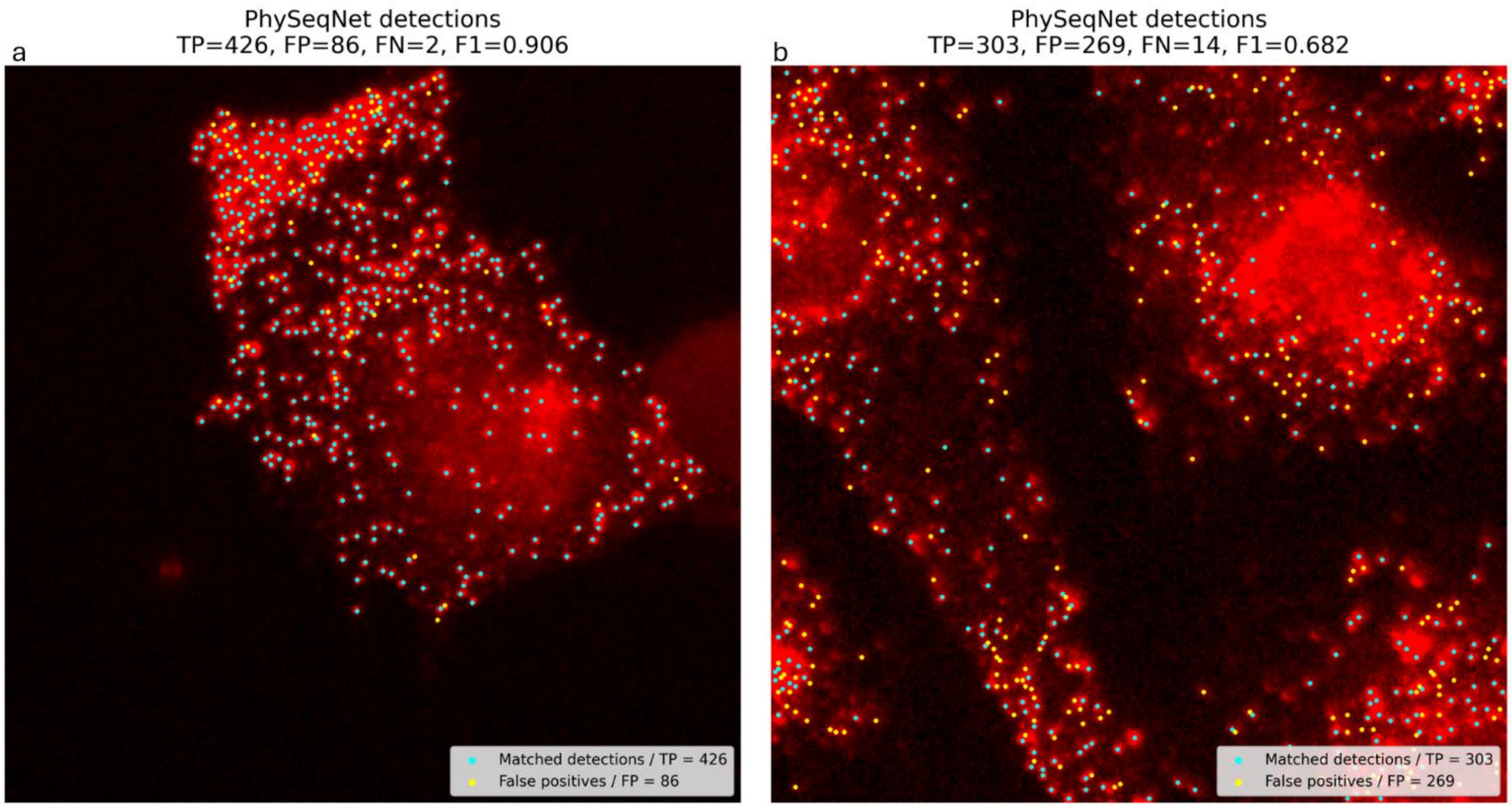

Figure S9. False-positive analysis of zero-shot PhySeqNet predictions on Piscis benchmark images [5]. PhySeqNet detections are classified relative to the reference annotations as matched detections/true positives (cyan) or false positives (yellow). a, PhySeqNet recovers 426 true-positive detections with 86 false positives and 2 false negatives (F1 = 0.906). b, In a more challenging image, PhySeqNet recovers 303 true positives with 269 false positives and 14 false negatives (F1 = 0.682). The small number of false negatives in both examples indicates high source-recovery sensitivity, while the reduced F1 score in the second image is driven primarily by excess detections under mismatched background conditions or dim spots not labeled as ground truth annotation.

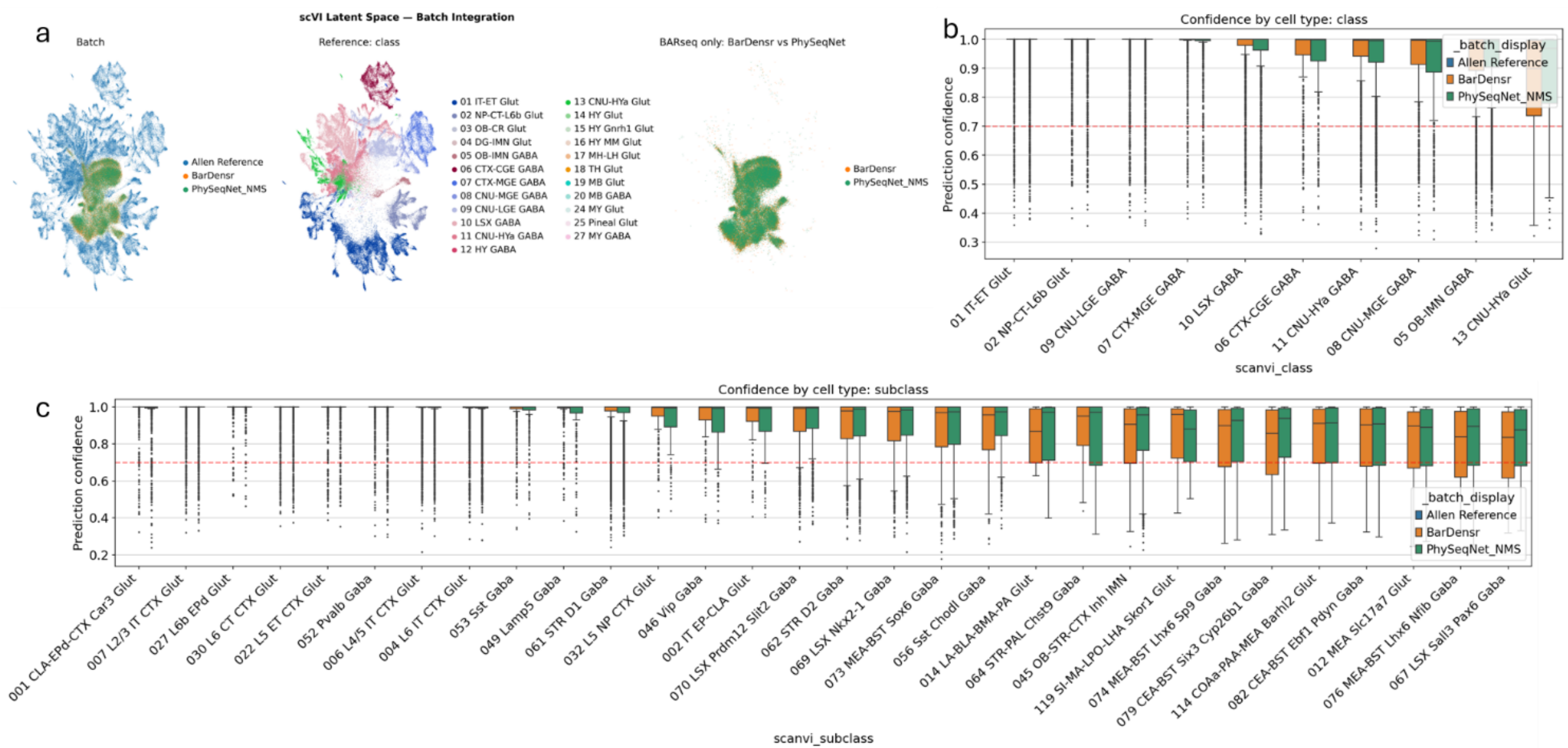

Figure S10. scVI/scANVI integration and cell-type prediction confidence for single-shot density estimation from BarDensr and PhySeqNet with NMSB. a, Joint scVI latent space showing integration of the Allen Brain Cell Atlas reference with BARseq cells processed using BarDensr or PhySeqNet with NMSB. The embedding is shown colored by dataset/batch, reference cell class, and BARseq processing method. b,c, scANVI prediction-confidence distributions for class- and subclass-level assignments, respectively, comparing Allen reference cells with BarDensr- and PhySeqNet-derived BARseq cells. The dashed horizontal line indicates the confidence threshold of 0.7 used for high-confidence assignments.

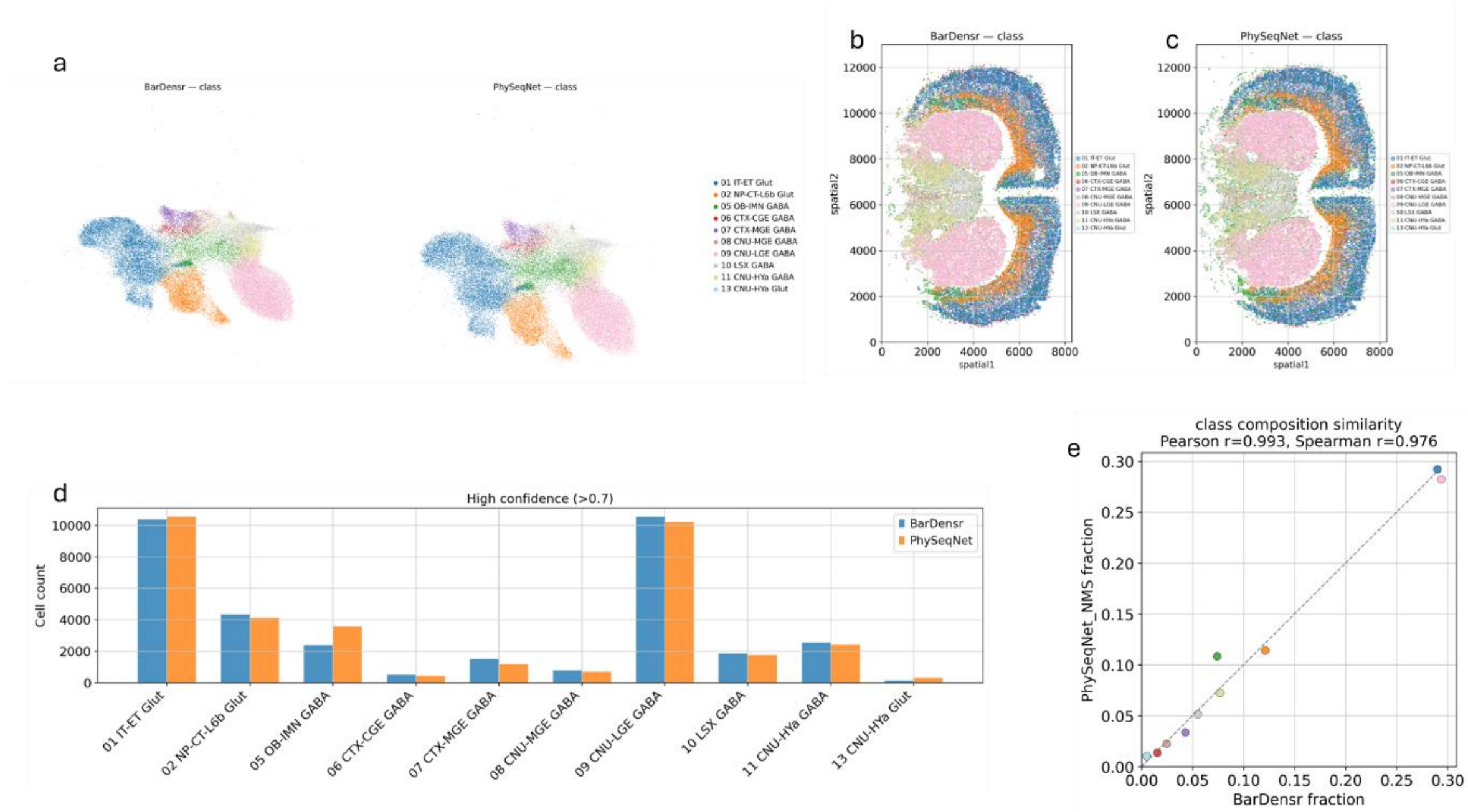

Figure S11. Class-level cell-type assignments are highly concordant between BarDensr and PhySeqNet. a, UMAP representations of BARseq cells colored by transferred scANVI class labels for BarDensr and PhySeqNet with NMSB. b,c, Spatial distributions of transferred class labels across the tissue section for the two pipelines. d, Numbers of high-confidence cells (scANVI confidence > 0.7) assigned to each major class. e, Comparison of class fractions between pipelines, showing strong agreement in cell-type composition (Pearson (r=0.993), Spearman (r=0.976)). The transcript recovery obtained with PhySeqNet therefore preserves broad class-level cellular organization.

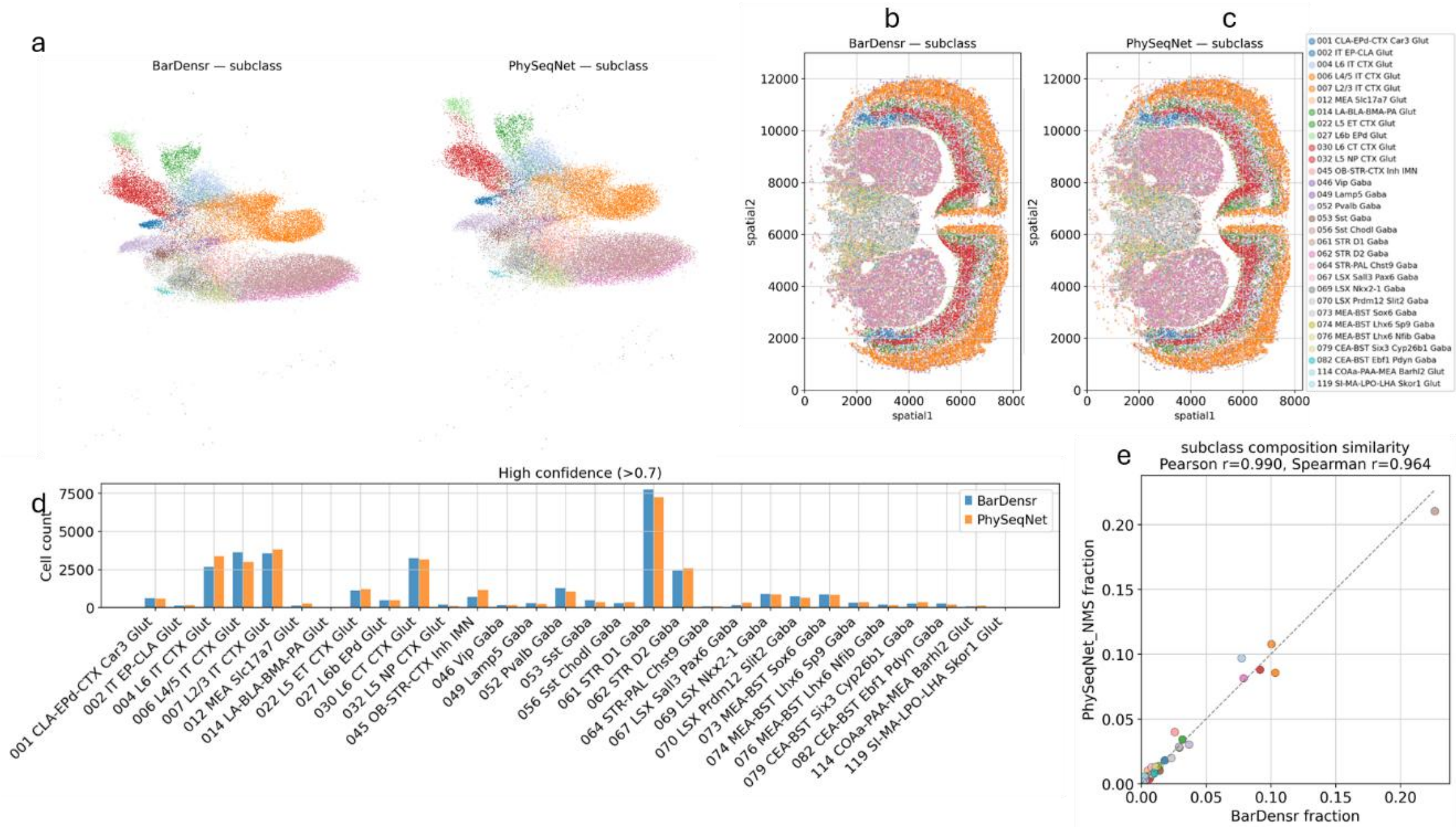

Figure S12. Subclass-level cell-type composition remains concordant between BarDensr and PhySeqNet with NMSB. a, UMAP representations colored by transferred scANVI subclass labels for BarDensr and PhySeqNet with NMSB. b,c, Spatial distributions of subclass assignments across the tissue section. d, Numbers of high-confidence cells (scANVI confidence > 0.7) assigned to each subclass. e, Subclass fractions are highly correlated between the two pipelines (Pearson (r=0.990), Spearman (r=0.964)), indicating that transcript recovery with PhySeqNet does not substantially alter the inferred subclass composition of the tissue.

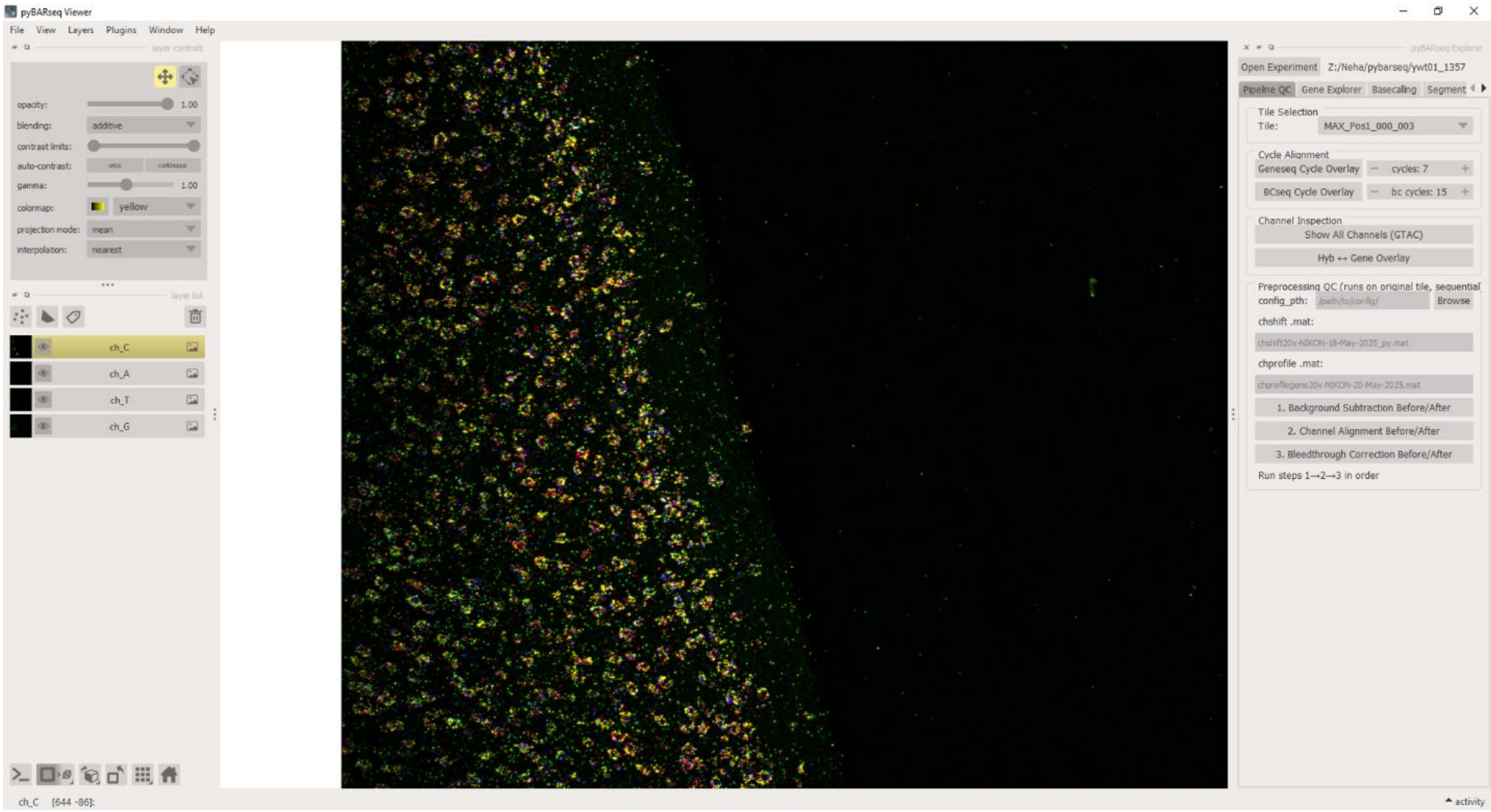

Figure S13. Interactive visualization and quality control using the pyBARseq napari viewer. Representative screenshot of the pyBARseq napari plugin showing multichannel BARseq fluorescence data together with image-layer controls and the docked pyBARseq analysis visualization interface. The viewer provides interactive access to tile and channel visualization and pipeline intermediate outputs, enabling inspection and quality control of pyBarSeq and downstream analysis within napari.